\documentclass[graybox]{svmult}
\usepackage{type1cm}
\usepackage{makeidx} 
\usepackage{graphicx}
\usepackage{multicol}
\usepackage[bottom]{footmisc}
\usepackage{newtxtext}
\usepackage[varvw]{newtxmath}
\makeindex
\usepackage{enumerate}
\usepackage[hyperindex,breaklinks]{hyperref}
\usepackage{tikz}
\usepackage{bbding,ifthen}
\usetikzlibrary{arrows,arrows.meta,shapes,positioning,calc,decorations.pathreplacing,calligraphy}
\tikzstyle{block} = [draw, rectangle,minimum height=3em, minimum width=3em]
\tikzstyle{sum} = [draw, circle]
\tikzstyle{box}=[rectangle, fill=gray!20, draw, minimum width=1.2cm, minimum height=0.5cm, align=center]

\usepackage[style=numeric-verb,
						backend=bibtex8,
						bibencoding=latin1,
						maxbibnames=99,doi=false,isbn=false,url=false,
						]{biblatex}
\bibliography{references}		
\usepackage{xurl}
\hypersetup{
	breaklinks=true,   
  pdfusetitle=true,  
        }		
						
\newcommand{\proofen}[1]{\begin{proof}#1
\end{proof}}

\newcommand{\aln}[1]{\noindent\begin{align*}#1\end{align*}}
\newcommand{\alnl}[1]{\noindent\begin{align}#1\end{align}}
\newcommand{\mat}[2]{\left(\begin{array}{#1}#2\end{array}\right)}

\newcommand{\V}{{\cal{V}}}

\renewcommand{\E}{{\cal{E}}}
\newcommand{\G}{\cal{G}} 
\renewcommand{\P}{{\cal{P}}}
\newcommand{\Z}{{\cal{Z}}}

\newcommand{\R}{ {\mathbb{R}} }
\newcommand{\N}{ {\mathbb{N}} }

\newcommand{\DeltaB}{\boldsymbol{\Delta}}

\newcommand{\norm}[1]{\lVert#1\rVert}
\newcommand{\norma}[1]{\left\lVert#1\right\rVert}
\newcommand{\tp}{^{\mathrm{T}}}
\newcommand{\inv}{^{-1}}
\newcommand{\im}{\text{ran}}
\newcommand{\kron}{\otimes}

\newcommand{\lbreak}{\textcolor[rgb]{1,0,0}{break}\\}
\renewcommand{\lbreak}{\\}

\begin{document}
\title*{Robust performance for switched systems with constrained switching and its application to weakly hard real-time control systems}
\titlerunning{Robust performance for switched systems with constrained switching}
\author{Simon Lang, Marc Seidel, and Frank Allgöwer}
\institute{Simon Lang \at University of Stuttgart, Institute for Systems Theory and Automatic Control,\\\email{simon.lang@ist.uni-stuttgart.de}
			\and Marc Seidel \at University of Stuttgart, Institute for Systems Theory and Automatic Control,\\ \email{marc.seidel@ist.uni-stuttgart.de}
			\and Frank Allgöwer \at University of Stuttgart, Institute for Systems Theory and Automatic Control,\\ \email{frank.allgower@ist.uni-stuttgart.de}}

\maketitle

\abstract*{
Many cyber-physical systems can naturally be formulated as switched systems with constrained switching. This includes systems where one of the signals in the feedback loop may be lost. Possible sources for losses are shared or unreliable communication media in networked control systems, or signals which are discarded, e.g., when using a shared computation device such as a processor in real-time control applications. The use of switched systems with constrained switching is not limited to cyber-physical systems but, includes many other relevant applications such as power systems and modeling virus mutations. In this chapter, we introduce a framework for analyzing and designing controllers which guarantee robust quadratic performance for switched systems with constrained switching. The possible switching sequences are described by the language of a labeled graph where the labels are linked to the different subsystems. The subsystems are allowed to have different input and output dimensions, and their state-space representations can be affected by a broad class of uncertainties in a rational way. The proposed framework exploits ideas from dissipativity-based linear control theory to derive analysis and synthesis inequalities given by linear matrix inequalities. We demonstrate how the proposed framework can be applied to the design of controllers for uncertain weakly hard real-time control systems -- a system class naturally appearing in networked and real-time control.
}

\abstract{
Many cyber-physical systems can naturally be formulated as switched systems with constrained switching. This includes systems where one of the signals in the feedback loop may be lost. Possible sources for losses are shared or unreliable communication media in networked control systems, or signals which are discarded, e.g., when using a shared computation device such as a processor in real-time control applications. The use of switched systems with constrained switching is not limited to cyber-physical systems but, includes many other relevant applications such as power systems and modeling virus mutations. In this chapter, we introduce a framework for analyzing and designing controllers which guarantee robust quadratic performance for switched systems with constrained switching. The possible switching sequences are described by the language of a labeled graph where the labels are linked to the different subsystems. The subsystems are allowed to have different input and output dimensions, and their state-space representations can be affected by a broad class of uncertainties in a rational way. The proposed framework exploits ideas from dissipativity-based linear control theory to derive analysis and synthesis inequalities given by linear matrix inequalities. We demonstrate how the proposed framework can be applied to the design of controllers for uncertain weakly hard real-time control systems -- a system class naturally appearing in networked and real-time control.
}
\section{INTRODUCTION}
Switched systems and their generalization to switched systems with constrained switching have turned into a useful system class for modeling real-world problems. Their use in networked control systems (NCSs)~\cite{Hespanha2007}, robotics~\cite{Martin2009}, and virus mutation dynamics~\cite{HernandezVargas2011} are only a few examples of possible applications. For motivating the utility of switched systems with constrained switching, we consider a problem in networked control in more detail in the sequel. In networked control, an important problem is to find a controller which stabilizes a given system and renders the closed-loop system such that it achieves a desired performance specification. In contrast to the classical control problem, the control inputs can be lost at some time instants due to network effects. Different approaches have been proposed to model the involved loss process. Apart from stochastic models like Bernoulli processes~\cite{Seiler2001}, deterministic descriptions have been used to describe the packet dropouts of the network leading to the loss of the control input.
Examples include bounded packet loss~\cite{Xiong2007} and, as a generalization thereof, weakly hard real-time (WHRT) constraints~\cite{Bernat2001}.
WHRT constraints are deterministic window-based descriptions of the loss process. The most common type of these constraints constitutes a lower bound on the amount of successful transmissions that can occur within a moving time window of a given length. For example, in any time window of length three, there are at least two packet transmissions successful. Guaranteeing such bounds might seem difficult but there exist scheduling strategies which provides these bounds routinely, e.g., in CAN systems~\cite{Broster2002}. Originally, WHRT constraints were used to schedule control tasks on a shared computation resource (e.g., a processor), in a real-time control setting.
Therein, in some invocations of the control task, the processor might not finish the computation of the control signal before its deadline.
This may happen, for example, because the computation resource is currently occupied by another task with higher priority.
Hence, similar to NCSs, the control signal can be considered as lost. Both practical scenarios can be described by the same system class - the so-called WHRT control systems. The advantage of deterministic models like WHRT control systems is that they can be used to provide deterministic guarantees. Such guarantees are especially relevant in safety-critical applications.

In both fields, real-time control and NCSs, there exist works dealing with stability properties of WHRT control systems.
For real-time control problems, stability investigations can be found in~\cite{Maggio2020,Vreman2022} and quite recently WHRT constraints also have been used to model the packet dropout in NCSs~\cite{Blind2015}. 
The WHRT control system can be equivalently represented by a switched system with constrained switching, where the possible switching sequences are described by the language of a labeled graph~\cite{Linsenmayer2017,Linsenmayer2021}.
The same method is applicable if the WHRT constraint describes the deadline miss process. The discussed example shows the practical relevance of switched systems with constraint switching for modeling real-world problems and especially for cyber-physical systems. Therefore, we consider switched systems with constrained switching in this chapter.
Apart from the reformulation of WHRT control systems into switched systems,~\cite{Linsenmayer2017,Linsenmayer2021} present results for designing stabilizing controllers.
The approach can, under mild assumption on the graph, be generalized to switched systems whose switching behaviors are described by languages of graphs.
These switched systems are also called constrained switched systems and we focus on this system class in this work due to their practical relevance. Note that switched systems with no constraints on the possible switching sequences are special instants of constrained switched systems.

In many practical applications, the control objective is not only to stabilize a system but to achieve a desired performance level. A commonly used family of performance measures is quadratic performance which includes, e.g., the worst-case energy amplification, also called the $\ell^2$-gain, and passivity. This motivates to develop a framework for the design of controllers for constrained switched systems such that the closed-loop system satisfies a desired quadratic performance property. The controller should not only achieve quadratic performance on the model which was used to design it but it should also achieve performance on the actual plant. The difference between the actual plant and the model used for controller synthesis can originate,  e.g., from an inaccurate or simplified system modeling. Such a difference is often described using uncertain models. In uncertain models, the uncertainty represents, e.g., unknown dynamics or parameters. Therefore, a practical relevant framework should also possess the possibility to design controllers which achieve quadratic performance in the face of uncertainties in the model. In this chapter, we propose such a framework. 

For the case of switched systems with arbitrary switching and without uncertainties and the $\ell^2$-gain as a performance measure, results for analysis and controller synthesis are presented in~\cite{Daafouz2002}. Similar conditions for constrained switched systems are proposed in~\cite{Falchetto2017}. In contrast to~\cite{Daafouz2002}, the authors of~\cite{Falchetto2017} only propose conditions for analysis with respect to the $\ell^2$-gain and an $\mathcal{H}^2$-like criterion for constrained switched systems. They use the obtained analysis results to calculate a set of allowed switching sequences such that a given switched system whose switching sequences are restricted to the calculated set satisfies a desired performance level. Moreover, the proposed solution does not allow for more than one edge connecting two nodes of the underlying graph. Results for the $\ell^2$-gain without this restriction can be found in~\cite{Philippe2016} where the conditions for the $\ell^2$-gain of arbitrary linear time-varying systems, presented in~\cite{Essick2014}, are exploited. The authors of~\cite{Philippe2016} aim to find a tight characterization of the $\ell^2$-gain of a constrained switched system and do not consider the problem of designing controllers. However, the presented results can be extended in this direction. In our preliminary conference paper~\cite{Seidel2023}, we present results for the special case of controller synthesis for WHRT control systems with an $\ell^2$-performance criterion. The results from~\cite{Seidel2023} can also be applied to arbitrary constrained switched systems. However, the approach proposed in~\cite{Seidel2023} relies on the results for switched systems from~\cite{Daafouz2002} and not on the previously mentioned publications~\cite{Philippe2016,Essick2014} which use results for arbitrary time-varying systems. All the cited literature has in common that they do not explicitly use a general mechanism to obtain the corresponding results. Therefore, there is no direct way to extend the results, which consider mostly the $\ell^2$-gain, to arbitrary quadratic performance criteria, which is one of the objectives of this work.

For linear systems, such a general mechanism is given by dissipativity. Since it was originally introduced by Willems~\cite{Willems1972}, dissipativity and its consequences have been a powerful tool for linear system theory. Its applications include the analysis and design of (robust) controllers with performance guarantees via convex optimization either model-based~\cite{Boyd1994,Scherer2015} or data driven~\cite{Waarde2022}. For continuous-time switched systems with arbitrary switching, the authors of~\cite{Zhao2008} introduce a notion of dissipativity and show how passivity and the $\ell^2$-gain can be incorporated in this framework. In the case of discrete-time unconstrained switched systems, the concept of dissipativity is also exploited to analyze interconnections of switched systems~\cite{McCourt2012} and to design switching laws which ensure that the switched system is dissipative~\cite{Jungers2019}. In~\cite{McCourt2012} and~\cite{Jungers2019}, switching quadratic supply rates are considered. Results for the design of controllers which guarantee dissipativity for discrete-time unconstrained switched systems with dwell time constraints can be found in~\cite{Wen2019}. Therein, the authors propose a definition of dissipativity which is close to the definition of quadratic performance and in this way also propose conditions for quadratic performance in terms of linear matrix inequalities (LMI). In contrast to~\cite{Zhao2008,McCourt2012,Jungers2019}, the approach presented in~\cite{Wen2019} is restricted to a global non-switching supply rate. Therefore, it implicitly assumes that the input and output dimensions of the constrained switched system are constant. For $\ell^2$-performance of WHRT control systems, we show in~\cite{Seidel2023} that it is beneficial to consider varying input and output dimensions. Therefore, the restriction on a global supply rate limits the practical applicability of such results. Moreover, most of the mentioned approaches neglect that constraints on the set of allowed switching sequences can originate from the considered problem itself as it is the case, for example, for WHRT control systems. This also implies that a restriction of the allowed switching sequences cannot be used to support controller synthesis. Therefore, the mentioned approaches described in~\cite{Zhao2008,McCourt2012,Jungers2019} cannot be used to design controllers for constrained switched systems as we consider it in this work.

Apart from the analysis and synthesis results for dissipativity, the authors of~\cite{Wen2019} also consider the case where the dynamic matrices of the state-space representations of the involved subsystems are subject to additive uncertainties. The possibility to consider uncertain state-space representations is an important property, but the assumption that the uncertainty is additive to the dynamic matrix might be restrictive for real-world applications.

In this work, we aim to close the mentioned gaps and propose a framework for controller synthesis for constrained switched systems such that the closed-loop system satisfies a desired quadratic performance criterion even in the case that the switched system is influenced by uncertainties.

\textbf{Contribution. }We propose a unifying framework based on dissipativity for the design of controllers for switched systems with constrained switching such that the closed-loop satisfies a given quadratic performance criterion. The considered switched system can be affected by uncertainties. Moreover, we illustrate how other performance criteria like the energy-to-peak gain can be incorporated into our framework. 

For describing the constraints on the possible switching sequences, we associate to each subsystem a label. The allowed switching sequences are then described by all sequences of labels which are in the language of a given labeled graph.  We consider supply rates which can be dependent on the labels and therefore our approach can be applied to constrained switched systems with different input and output dimensions of the subsystems. 

We show that with the dissipativity-based framework, arbitrary uncertain systems can be considered as long as linear fractional representations are given and the graph of the uncertainty renders a choosable quadratic form positive. 

We also demonstrate the practical relevance of the framework by applying it to the design of controllers for weakly hard real-time control systems such that the closed-loop system is guaranteed to achieve robust quadratic performance. In the literature, only our previous published work~\cite{Seidel2023} considers classical control performance guarantees for weakly hard real-time control systems and robustness properties were not considered for this class of systems so far. 
 
\textbf{Outline. }The remainder of this chapter is organized as follows. After a short paragraph on notation, we introduce the notions of constrained switched systems, dissipativity, and of quadratic performance in Section~\ref{sec:system_formulation}. In Section~\ref{sec:analysis}, we show how dissipativity implies quadratic performance for constrained switched systems and state conditions for dissipativity in terms of LMIs. First, the nominal case is discussed in Subsection~\ref{sec:subnominal} and then the ideas are extended to the case of uncertain systems in Subsection~\ref{sec:robustness}. We exploit the obtained results to derive convex conditions for controller synthesis in Section~\ref{sec:synthesis}. In Section~\ref{sec:WHRT}, we apply the results to weakly hard real-time control systems. The chapter is concluded by a numerical example in Section~\ref{sec:numerical}.

\textbf{Notation. }The sets of real numbers and non-negative integers are denoted by $\R$ and $\N$, respectively. The standard Euclidean norm of $\R^n$ is abbreviated by $\norm{\cdot}$ and the maximum norm by $\norm{\cdot}_{\infty}$. For a matrix $A$, the spectral norm is written as $\norm{A}$. The discrete-time system $(A,B,C,D)$ is described by the equations\aln{x(t+1)&=Ax(t)+Bu(t)\\ y(t)&=Cx(t)+Du(t)\\ x(0)&=0}for the time $t\in\N$, where $u$ and $y$ denote the input and output, respectively, and $x$ is called state of the system. If another initial condition of the state is considered, then it is explicitly stated. We denote the kernel and the image of a matrix $A$ by $\mathrm{ker}(A)$ and $\mathrm{ran}(A)$, respectively. For a Hermitian matrix $M$, we write $M\succ0~(M\prec0)$ if it is positive- (negative-) definite and $M\succeq 0~(M\preceq0)$ if it is positive- (negative-)semidefinite.  For matrices $A_1,A_2,...,A_N$, $\mathrm{col}(A_1,A_2,...,A_N)$ denotes the block column vector of them, $\mathrm{row}(A_1,A_2,...,A_N)$ the block row vector, and $\mathrm{diag}(A_1,A_2,...,A_N)$ the block diagonal matrix with diagonal blocks $A_1,A_2,...,A_N$. Elements which can be inferred by symmetricity are abbreviated by $\bullet$. Dimensions are omitted if they can be inferred from the context. The Kronecker product is denoted by $\otimes$. Finally, $I$ and $0$ denote the identity and zero matrix of suitable dimensions.

\section{SYSTEM FORMULATION}\label{sec:system_formulation}
In this section, we formally define the considered switched linear system with constrained switching which is also called constrained switching linear system (CSLS). In words, a CSLS is a system whose dynamic switches inside a finite set of linear dynamics and the switching behavior is constrained using a graph. 
Before we can state the formal definition of a CSLS, we need the notion of a constraining graph.
\begin{definition}[Constraining graph]
	The tuple $\G=(\V,\E)$ consisting of a given finite set of nodes $\V\subset\N$, and a finite set of labeled edges $\E\in\V\times\V\times\Z$ with $\Z=\{1,..,m\}$ for a given number of labels $m\in\N$ such that each node has an incoming and outgoing edge is called constraining graph.
\end{definition}

We remark that the cardinality of $\E$ can be different from the number of labels $m$, i.e., one label can be assigned to different edges. Subsequently, for a given labeled edge $e=(i,j,l)\in\E$, we denote by $s(e)=i$ and $f(e)=j$ the tail and head of the edge and by $\sigma(e)=l$ its label. 
Now we are able to define a CSLS.
\begin{definition}[CSLS]\label{def:CSLS}
	Given a constraining graph $\G$ with $m$ labels and a family of systems $\Theta=((A_k,B_k,C_k,D_k))_{k=1}^{m}$, a constrained switched linear system (CSLS) is described by the equations\aln{
		x(t+1)&=A_{\sigma(e(t))}x(t)+B_{\sigma(e(t))}w(t)\\
		z(t)&=C_{\sigma(e(t))}x(t)+D_{\sigma(e(t))}w(t)\\
		e(t+1)&\in\{{\tilde{e}}\in\E:s(\tilde{e})=f(e(t))\}
	}for $e(0)\in\E$ and $x(0)\in\R^n$, where $w$ denotes the input and $z$ the output signal.
\end{definition}

The last equation in the system description has to be interpreted in the way that $e(t+1)$ can be an arbitrary element in the set defined by $e(t)$. Observe that an unconstrained switched system can be considered as a special instance of Definition~\ref{def:CSLS}. The possible evolution of $e$, which is a sequence of edges, is fully described by the constraining graph. In this way, it also describes all possible sequences of labels. The sequence of labels is used to decide which system dynamic in $\Theta$ is considered at $t\in\N$ and therefore a given sequence of labels $\sigma(e)$ can be interpreted as a given sequence of state-space representations. The notation of $\sigma(e)$ has to be understood as a pointwise application of $\sigma$. For later purposes, it is essential to observe that we do not require that all systems in $\Theta$ have the same number of inputs and outputs and therefore the number of inputs and outputs of the CSLS can change over time. We only require that the systems in $\Theta$ have the same state-dimension.
The corresponding input and output dimensions to $\sigma(e)$ are collected in two sequences and are denoted by $d_\mathrm{i}(\sigma(e))$ and $d_\mathrm{o}(\sigma(e))$, respectively, i.e, the operators $d_\mathrm{i}$ and $d_\mathrm{o}$ map a given sequence of labels in the corresponding sequences of input and output dimensions. For the remainder of this work, we introduce the notion of the node of a CSLS at time $t\in\N$ which is the value $s(e(t))$. Observe that the nodes of a CSLS coincide with the nodes of the constraining graph.

In the sequel, we are mainly interested in stability and quadratic performance of CSLSs. The definition of asymptotic stability can be easily carried over from its classical definition as introduced in textbooks, e.g.,~\cite{Khalil2002}. For the sake of completeness, it is stated next.
\begin{definition}[Stability]
	The CSLS is called stable if for all $\epsilon>0$ there exists a $\delta>0$ such that\aln{\norm{x(0)}<\delta \Rightarrow \norm{x(t)}<\epsilon}for all $t\geq 0$, $e(0)\in\E$, $w=0$, and all possible evolutions of $x$ and $e$.
\end{definition}
\begin{definition}[Asymptotic Stability]
	The CSLS is called asymptotically stable if it is stable and there exists $\tilde{\delta}>0$ such that\aln{\norm{x(0)}<\tilde{\delta} \Rightarrow \lim\limits_{t\to\infty}\norm{x(t)}=0}for all $t\geq 0$, $e(0)\in\E$, $w=0$, and all possible evolutions of $x$ and $e$.
\end{definition}

Before we are able to define quadratic performance, we have to define a proper space for the input signals $w$. Recall that the input dimensions of the state-space representations in $\Theta$ can be different and therefore the classical space of all square summable functions $\ell^2$ cannot be used. For a bounded sequence $N=(n(k))_{k=0}^{\infty}$ in $\N\setminus\{0\}$, we define\aln{\ell^2_N=\big\{\{w(t)\}_{t\in\N}:w(t)=\R^{n(t)},t\in\N\text{ and }\sum \limits_{k=0}^{\infty}w(k)\tp w(k)<\infty \big\}} and $\norm{w}_{\ell^2_N}=\sum \limits_{k=0}^{\infty}w(k)\tp w(k)$ as a norm on $\ell^2_N$. It is not hard to see that for a fixed sequence $N$ in $\N\setminus\{0\}$ with bounded values, $\ell^2$ (equipped with the usual norm) is isometrically isomorphic to $\ell^2_N$ (equipped with $\norm{\cdot}_{\ell^2_N}$). We denote this isometry by $q_N$. 

Now we are able to define quadratic performance of a CSLS.
\begin{definition}\label{def:per}
	The CSLS satisfies quadratic performance with the label-dependent and symmetric performance index\aln{P^p=(P_l^p)_{l=1}^m=\mat{cc}{Q_l^p&S_l^p\\(S_l^p)\tp&R_l^p}}where $P_l^p, l\in\Z$ is symmetric if it is asymptotically stable and there exists an $\tilde{\epsilon}>0$ such that\aln{\sum \limits_{k=0}^{\infty}\mat{c}{w(k)\\z(k)}\tp\mat{cc}{Q^p_{\sigma(e(k))}&S_{\sigma(e(k))}^p\\\left(S_{\sigma(e(k))}^p\right)\tp&R_{\sigma(e(k))}^p}\mat{c}{w(k)\\z(k)}\leq-\tilde{\epsilon}\norm{w}_{\ell_{d_\mathrm{i}(\sigma(e))}^2}} for $x(0)=0$, $e(0)\in\E$, $w\in l_{d_\mathrm{i}(\sigma(e))}^2$ and all possible evolutions of $x$ and $e$.
\end{definition}

If all involved state-space representations have the same input, then $l_{q_\mathrm{i}(\sigma(e))}^2=\ell^2$. If furthermore a non-switching performance index, i.e., $P_l=\hat{P},~l\in\Z$ is considered and all state-space representations have also the same output dimension, then the classical definition of quadratic performance is obtained. Observe that if we choose $Q_l=-\gamma^2 I$, $S_l=0$, and $R_l=I$, we obtain the classical characterization of $\ell^2$-performance with gain smaller or equal to $\gamma$ as we considered it in~\cite{Seidel2023}.
\section{ANALYSIS}\label{sec:analysis}
In this section, we derive conditions in terms of LMIs which guarantee that the CSLS defined by Definition~\ref{def:CSLS} is asymptotically stable and satisfies quadratic performance for a given performance index $P^p$. In the first subsection, we consider nominal CSLSs and extend the obtained results to uncertain CSLSs which are defined by a family of state-space representations affected by uncertainties in the second subsection.

\subsection{Nominal CSLSs}\label{sec:subnominal}
In order to derive the desired conditions for quadratic performance of a nominal CSLS, we pursue the following steps. We first state a condition for asymptotic stability (Theorem~\ref{theo:as}) and then use the concept of dissipativity to derive an additional LMI for every labeled edge which ensures quadratic performance of the CSLS (Theorem~\ref{theo:diss2P}). The derived two LMIs for each labeled edge are combined in Theorem~\ref{thm:main} which also provides equivalent reformulations. These reformulations are used to design controllers for CSLSs in Section~\ref{sec:synthesis}. Under some assumptions, further equivalent conditions are provided by Corollary~\ref{thm:mainC} and are exploited for the design of robust controllers for CSLSs in Section~\ref{sec:synthesis}. The subsection is concluded by an application of the introduced mechanism to the problem of finding analysis conditions for CSLSs which guarantee a bound on the energy-to-peak amplification of a CSLS (Lemma~\ref{thm:en2Peak}).

We start our discussion with a sufficient condition for asymptotic stability of a CSLS which is well-known in the literature.
\begin{lemma}\label{theo:as}
	The CSLS defined by Definition~\ref{def:CSLS} is asymptotically stable if there exist $X_i \succ 0$ for all $i\in\V$ such that\alnl{\label{eq:lyap}A_l\tp X_j A_l-X_i\prec 0} for all $(i,j,l)\in\E$.
\end{lemma}
\proofen{A proof can be found in~\cite{Linsenmayer2017}. It employs classical Lyapunov theory arguments in combination with node dependent Lyapunov functions.}
Observe that~\eqref{eq:lyap} is already an LMI. For deriving conditions for quadratic performance, we employ ideas from dissipativity theory. The authors of~\cite{Wen2019} pointed out how dissipativity properties of an unconstrained switched system can be guaranteed. In contrast to the cited work, we show the same for CSLSs and also use a definition of dissipativity which is closer to the original one by Willems~\cite{Willems1972}. Similar to the condition of asymptotic stability (Lemma~\ref{theo:as}), we allow the storage function to be node dependent. For ease of notation, we restrict ourselves to quadratic storage functions and supply rates. This choice can be motivated due to the numerically tractable conditions which can be inferred from it. We define the label-dependent supply rate\alnl{\label{eq:supplyRate}s^d_l(w(k),z(k))=\mat{c}{w(k)\\z(k)}\tp P_l\mat{cc}{w(k)\\z(k)},\qquad l\in\Z} for the family of symmetric matrices $P=(P_l)_{l=1}^{m}$ which we call dissipativity index. Similar to the performance index, we partition the dissipativity index according to the input and output dimension and introduce the notation\aln{P_l=\mat{cc}{Q_l&S_l\\S_l\tp&R_l},\quad l\in\Z.}With this notation, we can introduce dissipativity.
\begin{definition}[Dissipativity]\label{def:diss}
	The CSLS is dissipative with a label-dependent quadratic storage function and supply rate $(s_l^d)_{l=1}^{m}$ defined in \eqref{eq:supplyRate}, if there exist $X_i=X_i\tp$, $i\in\V$ and $\epsilon>0$ such that\alnl{\label{eq:diss}x(t+1)\tp X_{j}  x(t+1) + \mat{c}{w(t)\\z(t)}\tp P_{l} \mat{c}{w(t)\\z(t)}\leq x(t)X_{i}x(t) -\epsilon w(t)\tp w(t)}for all $x(t)\in \R^n$, $w(t)\in \R^{d_\mathrm{i}(l)}$, $z(t)=C_{l}x(t)+D_lw(t)$,  $t\geq 0$, and $(i,l,j)\in\E$.
\end{definition}

In our definition of dissipativity we only consider one time step in the evolution of the state of the CSLS. However, by applying the definition iteratively with respect to the evolution of $e$, we can guarantee a similar inequality for multiple time steps. This observation is the key mechanism in the proof of the next theorem which links dissipativity of a CSLS to its performance properties.
\begin{theorem}\label{theo:diss2P}
	If the CSLS is asymptotically stable and dissipative with the label-dependent supply rate induced by the index $P$, then it achieves quadratic performance with performance index $P^p=P$.
\end{theorem}
\proofen{For an arbitrary and admissible evolution of $e$ and $h>0$, iterative concatenation of~\eqref{eq:diss} according to the switching sequence defined by $e$ results in\aln{x(h)\tp X_{f(e(h-1))}  x(h) + \sum \limits_{k=0}^{h-1}\mat{c}{w(k)\\z(k)}\tp P_{\sigma(e(k))} \mat{c}{w(k)\\z(k)}\leq -\epsilon\sum \limits_{k=0}^{h-1} w(k)\tp w(k)}for $x(0)=0$ and $w\in l_{d_\mathrm{i}(\sigma(e))}^2$.\newpage Because $h>0$ arbitrary and the CSLS is asymptotically stable, we infer\aln{\sum \limits_{k=0}^{\infty}\mat{c}{w(k)\\z(k)}\tp P_{\sigma(e(k))} \mat{c}{w(k)\\z(k)}\leq -\epsilon\sum \limits_{k=0}^{\infty} w(k)\tp w(k)}for all $w\in l_{d_\mathrm{i}(\sigma(e))}^2$ and possible evolutions of $x$ and $e$ which is the definition of quadratic performance (Definition~\ref{def:per}) with label-dependent index $P^p=P$.
In the last step we used the fact that $\norm{x}_{\ell^2}\leq c, c\in\R$ and therefore also $\lim\limits_{k\to\infty}\norm{x(k)}=0$. The validity of this statement is shown next. Asymptotic stability implies the existence of a $K\geq 1$ and $0<\rho<1$ such that\aln{\norm{A_{\sigma(e(t))}\cdots A_{\sigma(e(0))}}\leq K\rho^{t+1}}for all $t\geq 0$ and admissible sequences of edges $e$~\cite{Dai2012}. For a fixed sequence of edges $e$ and the corresponding state-sequence $x$ we infer\aln{
\norm{x}^2&=\sum\limits_{t=0}^{\infty}\norma{\sum\limits_{k=0}^{t-2}A_{\sigma(e(t-1))}\cdots A_{\sigma(e(k+1))}B_{\sigma(e(k))}w(k) +B_{\sigma(e(t-1))}w(t)}^2\\
&\leq c\sum\limits_{t=0}^{\infty}\sum\limits_{k=0}^{t-2}\norm{A_{\sigma(e(t-1))}\cdots A_{\sigma(e(k+1))}}^2\norm{w(k)}^2 +\norm{w(t)}^2\\
&\leq \frac{cK^2}{\rho^2}\sum\limits_{t=0}^{\infty}\sum\limits_{k=0}^{t-1}\left(\rho^{(t-k)}\norm{w(k)}\right)^2}with $c=\max\limits_{l\in\Z}\norm{B_l}$. An application of Young's convolution inequality shows the fact.}The inequality in Definition~\ref{def:diss} can be reformulated into an LMI. Hence, Lemma~\ref{theo:as} and Theorem~\ref{theo:diss2P} together provide sufficient conditions in terms of two LMIs for each labeled edge for quadratic performance of a CSLS. In the case that $R_l$ is positive semi-definite for all $l\in\Z$, both LMIs can be combined and also meaningfully reformulated. The reformulated LMIs will become crucial for controller synthesis later in this work. These reformulations are provided by the following theorem. \newpage
\begin{theorem}\label{thm:main}
Given a CSLS and an index $P=(P_l)_{l=1}^{m}$ with $R_l\succeq0$, $l\in\Z$ and the decomposition $R_l=U_l\tp\tilde{R_l}\inv U_l$ with $\tilde{R}\inv_l\succ0$ (such a decomposition exists due to $R_l\succeq0$), then there exist $X_i$ for all $i\in\V$ such that\aln{A_l\tp X_j A_l-X_i\prec 0} for all $(i,j,l)\in\E$ and the CSLS is dissipative with index $P$ if one of the following equivalent statements is satisfied:
\begin{enumerate}[(i)]
	\item \label{eq:char2} There exist $X_i\succ0$ for all $i\in\V$ such that\alnl{\label{eq:main}(\bullet)\tp\mat{cc}{-X_i&0\\0&X_j}\mat{cc}{I&0\\A_l&B_l}+(\bullet)\tp\mat{cc}{Q_l&S_l\\S_l\tp&R_l}\mat{cc}{0&I\\C_l&D_l}\prec0}for all $(i,j,l)\in\E$.
	\item \label{eq:char3} There exist $\tilde{X}_i=\tilde{X}_i\tp$ for all $i\in\V$ such that\alnl{\label{eq:mainS}\mat{cccc}{\tilde{X}_j&A_l\tilde{X}_i&B_l&0\\
																																					 \bullet&\tilde{X}_i&-\tilde{X}_iC_l\tp S_l\tp&\tilde{X}_iC_l\tp U_l\tp\\
																																					 \bullet&\bullet&-Q_l-S_lD_l-D_l\tp S_l\tp&D_l\tp U_l\tp\\
																																				   \bullet&\bullet&\bullet&\tilde{R}_l}\succ 0}for all $(i,j,l)\in\E$.																																				
	\item \label{eq:char4} There exist $\tilde{X}_i=\tilde{X}_i\tp$ and $G_i$ for all $i\in\V$ such that\alnl{\label{eq:mainP}\mat{cccc}{\tilde{X}_j&A_lG_i&B_l&0\\
																																									 \bullet&G_i+G_i\tp-\tilde{X}_i&-G_i\tp C_l\tp S_l\tp&G_i\tp C_l\tp U_l\tp\\
																																									 \bullet&\bullet&-Q_l-S_lD_l-D_l\tp S_l\tp&D_l\tp U_l\tp\\
																																									 \bullet&\bullet&\bullet&\tilde{R}_l}\succ 0}for all $(i,j,l)\in\E$.
\end{enumerate}
\end{theorem}
\proofen{We first show that Condition~\eqref{eq:char2} is equivalent to dissipativity and implies asymptotic stability. Dissipativity with index $P_l$ is equivalent to the existence of $X_i=X_i\tp$ for all $i\in \V$ such that~\eqref{eq:main} is satisfied. The left upper block is given by $A_l\tp X_jA_l-X_i+C_l\tp R_lC_l\prec 0$ which implies $A_l\tp X_jA_l-X_i\prec 0$ by assumption. Restricting $X_i\succ 0$ for $i\in\V$ completes the proof of Condition~\eqref{eq:char2}.

$\eqref{eq:char2} \Leftrightarrow\eqref{eq:char3}$: This is a simple application of a Schur complement argument~\cite{Horn1991} in combination with a congruence transformation with $\mathrm{diag}(X_j\inv,X_i\inv,0,0)$ and the substitution $\tilde{X}_i=X\inv_i$, $i\in\V$ for each fixed $(i,j,l)\in\E$.

$\eqref{eq:char3} \Leftrightarrow\eqref{eq:char4}$: Let $(i,j,l)\in\E$ be fixed. We will first show that $\eqref{eq:mainP} \Rightarrow\eqref{eq:mainS}$. To this end, note that the left side of~\eqref{eq:mainP} can be rewritten to $Q+U\tp G V+ V\tp G\tp U$ with\aln{
Q=\mat{cccc}{\tilde{X}_j&0&B_l&0\\ \bullet&-\tilde{X}_i&0&0\\ \bullet&\bullet&-Q_l-S_lD_l-D_l\tp S_l\tp&D_l\tp U_l\tp\\ \bullet&\bullet&\bullet&\tilde{R}_l},\quad
U\tp=\mat{c}{A_l\\I\\-S_lC_l\\U_lC_l},\quad
V=\mat{cccc}{0&I&0&0}.}If~\eqref{eq:mainP} holds, then a congruence transformation with a basis matrix of the kernel of $U$, denoted by $U_{\bot}$, yields $~{U\tp_{\bot}QU_{\bot}\succ 0}$. Elementary calculation and a Schur complement argument~\cite{Horn1991} show that this is equivalent to~\eqref{eq:mainS}. For the other direction, we can define $G_i=\tilde{X}_i$ which yields~\eqref{eq:mainP}. Because this holds for all $(i,j,l)\in\E$ we showed $\eqref{eq:char3} \Leftrightarrow\eqref{eq:char4}$. 
}We stress again that the combination of Theorem~\ref{theo:diss2P} and~\ref{thm:main} implies quadratic performance of the CSLS. Note that the dimensions of the inequalities in Condition~\eqref{eq:char3} and~\eqref{eq:char4} can depend on $(i,j,l)$ because we allow label-dependent indices where the involved matrices can have different dimensions.
In the special case that every matrix in the dissipativity index is invertible, further equivalent conditions can be stated. These reformulations will become crucial for the design of robust controllers. For an element $P_l$ in the index $(P_l)_{l=1}^m$, we partition its inverse according to the block structure of $P_l$ and write for the new blocks\aln{\mat{cc}{\tilde{Q}_l&\tilde{S}_l\\\tilde{S}_l\tp&\tilde{R}_l}=\mat{cc}{Q_l&S_l\\S_l\tp&R_l}\inv\text{ for all } l\in\Z.}
With this notation, the equivalent conditions are provided by the following corollary.
\begin{corollary} \label{thm:mainC}
	Under the assumptions of Theorem~\ref{thm:main} and the additional assumption that every element of $P$ is invertible, then the conditions~\eqref{eq:char2}-\eqref{eq:char4} in Theorem~\ref{thm:main} are equivalent to the following equivalent conditions:
	\begin{enumerate}[(i)]
		\item \label{eq:char2D} There exist $\tilde{X}_i\succ0$ for all $i\in\V$ such that\alnl{\label{eq:mainD}(\bullet)\tp\mat{cc}{\tilde{X}_j&0\\0&-\tilde{X}_i}\mat{cc}{I&0\\A_l\tp&C_l\tp}+(\bullet)\tp\mat{cc}{\tilde{R}_l&-\tilde{S}_l\tp\\-\tilde{S}_l&\tilde{Q}_l}\mat{cc}{0&I\\B_l\tp&D_l\tp}\succ 0}for all $(i,j,l)\in\E$ and $\tilde{Q_l}\preceq0$ for all $l\in\Z$.
		\item \label{eq:char3D} There exist $\tilde{X}_i=\tilde{X}_i\tp$ for all $i\in\V$ such that\alnl{\label{eq:mainSD}
			\mat{ccc}{\tilde{X}_i&\tilde{X}_iA_l\tp&\tilde{X}_iC_l\tp\\
				\bullet&\tilde{X}_j+B_l\tilde{Q}B_l\tp&-B_l\tilde{S}_l+B_l\tilde{Q}_lD_l\tp\\
				\bullet&\bullet&\tilde{R}_l-\tilde{S}_lD_l-D_l\tp\tilde{S}_l\tp+D_l\tilde{Q}_lD_l\tp}\succ 0}for all $(i,j,l)\in\E$ and $\tilde{Q_l}\preceq0$ for all $l\in\Z$.
		\item \label{eq:char4D} There exist $\tilde{X}_i=\tilde{X}_i\tp$ and $G_i$ for all $i\in\V$ such that\alnl{\label{eq:mainPD}
			\mat{ccc}{G_i+G_i\tp-\tilde{X}_i&G_iA_l\tp&G_iC_l\tp\\
				\bullet&\tilde{X}_j+B_l\tilde{Q}B_l\tp&-B_l\tilde{S}_l+B_l\tilde{Q}_lD_l\tp\\
				\bullet&\bullet&\tilde{R}_l-\tilde{S}_lD_l-D_l\tp\tilde{S}_l\tp+D_l\tilde{Q}_lD_l\tp}\succ 0}for all $(i,j,l)\in\E$ and $\tilde{Q_l}\preceq0$ for all $l\in\Z$.	
		
	\end{enumerate}
\end{corollary}
\proofen{
	The equivalence of the first condition of Corollary~\ref{thm:mainC} and Condition~\eqref{eq:char2} of Theorem~\ref{thm:main} is an application of the well-known Dualization lemma~\cite{Scherer2015}. Moreover, the substitution $\tilde{X}_i=X_i\inv$ for $i\in\V$ is applied.
	
	$\eqref{eq:char2} \Leftrightarrow\eqref{eq:char3}$: Similar to the proof of Theorem~\ref{thm:main}, this is an application of a Schur complement argument~\cite{Horn1991}.
	
	$\eqref{eq:char3} \Leftrightarrow\eqref{eq:char4}$: It follows the same lines of reasoning as in the proof of Theorem~\ref{thm:main}. The only difference is that the left side of~\eqref{eq:mainPD} is rewritten to $Q+U\tp G V+ V\tp G\tp U$ with\aln{
		Q&=\mat{ccc}{-\tilde{X}_i&0&0\\ \bullet&\tilde{X}_j+B_l\tilde{Q}B_l\tp&-B_l\tilde{S}_l+B_l\tilde{Q}_lD_l\tp\\\bullet&\bullet&\tilde{R}_l-\tilde{S}_lD_l-D_l\tp\tilde{S}_l\tp+D_l\tilde{Q}_lD_l\tp},\\
		U\tp&=\mat{c}{I\\0\\0},\quad
		V=\mat{ccc}{I&A_l\tp&C_l\tp}.}
		\phantom{text}
}
Before we continue, let us summarize what we have achieved so far: Lemma~\ref{theo:as} together with Theorem~\ref{theo:diss2P} provide conditions in terms of a set of LMIs for each labeled edge which guarantee that a given CSLS satisfies quadratic performance. These conditions are combined in one LMI for each edge and are reformulated in Theorem~\ref{thm:main} and Corollary~\ref{thm:mainC}. However, the presented framework is much more flexible because it can handle performance criteria other than pure quadratic performance. For example, the presented framework can also cope with performance criteria described by dissipativity inequalities and invariance constraints. We illustrate the offered flexibility of the presented dissipativity-based approach by stating a condition for the energy-to-peak gain, often called generalized $\mathcal{H}^2$-norm, in the next lemma.
\begin{lemma}
\label{thm:en2Peak}The CSLS with $D_l=0,~l\in\Z$ is asymptotically stable and satisfies\aln{\sup\limits_{0<\norm{w}_{l_{d_\mathrm{i}(\sigma(e))}^2}<\infty}\sup\limits_{t\geq 0}\frac{\norm{y(t)}}{\norm{w}_{l_{d_\mathrm{i}(\sigma(e))}^2}}<\gamma} for all possible evolutions of $x$ and $e$ with $x(0)=0$ and $e(0)\in\E$ if one of the following equivalent conditions holds:
\begin{enumerate}[(i)]
	\item\label{eq:H21}There exist $\tilde{X}_i=\tilde{X}_i\tp$ for all $i\in\V$ such that\aln{\mat{ccc}{\tilde{X}_j&A_l\tilde{X}_i&B_l\\\bullet&\tilde{X}_i&0\\\bullet&\bullet&\gamma I}\succ0, \quad \mat{cc}{\tilde{X}_i&\tilde{X}_iC_l\tp\\C_l\tilde{X}_i&\gamma I}\succ0}for all $(i,j,l)\in\E$.
	\newpage
	\item\label{eq:H22} There exist $\tilde{X}_i=\tilde{X}_i\tp$, $G_i$ with $i\in\V$ such that\aln{\mat{ccc}{\tilde{X}_j&A_lG_i&B_l\\\bullet&G_i+G_i\tp-\tilde{X}_i&0\\\bullet&\bullet&\gamma I}\succ0, \quad \mat{cc}{G_i+G_i\tp-\tilde{X}_i&G_i\tp C_l\tp\\C_lG_i&\gamma I}\succ0}for all $(i,j,l)\in\E$.
\end{enumerate}
\end{lemma}
\proofen{Similar to the proof of Theorem~\ref{thm:main} and Lemma~\ref{theo:as} the first inequality in~\eqref{eq:H21} implies asymptotic stability of the CSLS and validity of\aln{
x(T)\tp X_{f(e(T-1))}x(T)\leq \gamma\norm{w}_{l_{d_\mathrm{i}(\sigma(e))}^2}^2} for all $T\geq0$, $w\in l_{d_\mathrm{i}(\sigma(e))}^2$ and all admissible evolutions of $e$ and $x$ for all $e(0)\in\E$ and $x(0)=0$. We use again the abbreviation $X_i=\tilde{X}_i\inv$ for $i\in\V$. The second inequality in~\eqref{eq:H22} implies\aln{
\frac{1}{\gamma}\norm{z(T)}^2<x(T)\tp X_{i}x(T)} for all $i\in\V$ and all labels which are assigned to edges which start at $i$. This means that we can combine both inequalities and get\aln{\frac{1}{\gamma}\norm{z(T)}^2<x(T)\tp X_{f(e(T-1))}x(T)\leq \gamma\norm{w}_{l_{d_\mathrm{i}(\sigma(e))}^2}^2.}Because this holds for all $T\geq0$, $w\in l_{d_\mathrm{i}(\sigma(e))}^2$, and all admissible evolutions of $e$ and $x$ for all $e(0)\in\E$ and $x(0)=0$, the claim about the energy-to-peak gain follows. The equivalence of the first inequality in Condition~\eqref{eq:H21} and~\eqref{eq:H22} is shown similar to the proof of the equivalence of the Conditions~\eqref{eq:char3} and~\eqref{eq:char4} in Theorem~\ref{thm:main}. For showing the second inequality, we can define $G_i=\tilde{X_i}$ for all $i\in\V$ (this is the same substitution as we used it to show that the first inequality in Condition~\eqref{eq:H21} implies the first inequality in Condition~\eqref{eq:H22}). For the other direction, note that the inequalities in Condition~\eqref{eq:H22} imply that $G_i$ is invertible for $i\in\V$ and that $\tilde{X}_i\succ0$ for $i\in\V$. Hence, $G_i\tp\tilde{X}_i\inv G_i\succeq G_i+G_i\tp-\tilde{X}_i$ for all $i\in\V$. Substituting this in the second inequality and performing a congruence transformation with $\tilde{X_i}(G_i\tp)\inv$ on the first row/column yields the second inequality in Condition~\eqref{eq:H21}.}
\subsection{Uncertain CSLSs}\label{sec:robustness}
Apart from conditions for performance, the presented dissipativity-based approach can also be used to derive conditions which ensure robust stability and performance of a CSLS as we show in this section.
We assume that the CSLS is uncertain, i.e., the family of systems involved in Definition~\ref{def:CSLS} consists of uncertain state-space representations $\Theta=((\tilde{A}_l({\Delta}_l),\tilde{B}_l({\Delta}_l),\tilde{C}_l({\Delta}_l),\tilde{D}_l({\Delta}_l)))_{l=1}^m$ with a label-dependent uncertainty ${\Delta}_l$ which is an element of a given uncertainty set ${\DeltaB}_l, l\in\Z$. One way to think about uncertain CSLSs is to not only consider one CSLS, but consider the entire set of CSLSs which are parametrized by all possible uncertainties. With this interpretation, robust asymptotic stability and performance can be interpreted as nominal asymptotic stability and performance for all CSLSs in the set. It is simple to see that the robust counterparts of Lemma~\ref{theo:as} and Theorem~\ref{theo:diss2P} can be obtained if it is required that the conditions have to hold for all possible uncertainties. In robust control theory, it is common to represent an uncertain system as a feedback interconnection of a known system and an uncertainty~\cite{Zhou1996}. This representation is called linear fractional representation (LFR) and we assume that the uncertain state-space system representations are given in such a structure, i.e., for every $l\in\Z$ the state-space representation $(\tilde{A}_l({\Delta}_l),\tilde{B}_l({\Delta}_l),{C}_l({\Delta}_l),\tilde{D}_l({\Delta}_l))$ can be expressed by the set of equations
\begin{alignat*}{3}
	x(t+1)&=A_lx(t) &&+B^{w_u}_{l}w_u(t)&&+B^{w_p}_{l}w_p(t)\\
	z_u(t)&=C^{z_u}_{l}x(t) &&+D^{z_uw_u}_{l}w_u(t)&&+D^{z_uw_p}_{l}w_p(t)\\
	z_p(t)&=C^{z_p}_{l}x(t) &&+D^{z_pw_u}_{l}w_u(t)&&+D^{z_pw_p}_{l}w_p(t)\\
	w_u(t)&=\Delta_{l}z_u(t), && \quad \Delta_l\in\DeltaB_l
\end{alignat*}
for uncertainty sets $\DeltaB_l$, $l\in\Z$. The signals $w_{p}$ and $z_p$ denote the performance input and output, respectively. Clearly, this representation makes only sense if $I-D^{z_uw_u}_{l}\Delta_l$ is invertible for all $\Delta_l\in\DeltaB_l$. If the latter condition holds for all $l\in\Z$, we say that the LFR of the CSLS is well-posed. Because we do not assume that $D^{z_uw_u}_{l}$ is a zero matrix, we can consider uncertain systems where the dependence on the uncertainty is rational. Especially, we do not restrict ourselves to the case where the uncertainty affects the system description in an affine way. For the label-dependent uncertainty set $\DeltaB_l$ we only require that for each label $l\in\Z$ there exists a set of matrices $\P_l^{\Delta}$ such that the induced quadratic forms restricted on the graph of every element in the uncertainty set is positive semi-definite, i.e., \alnl{\label{eq:SepUn}\mat{c}{\Delta_l\\I}\tp P_l^{\Delta}\mat{c}{\Delta_l\\I}\succeq 0,\qquad \Delta_l\in\DeltaB_l}for all $P_l^{\Delta}\in\P_l^{\Delta}$ and for all $l\in\Z$. In the following, we will call $(P_l^{\Delta})_{l=1}^m$ a multiplier index. Such a representation of the uncertainty set is very common in literature and covers many different cases. For example, the case of real-repeated uncertainties or full-block uncertainties bounded in their norms can be easily captured by this representation. Recall that different multipliers can be diagonally combined to derive a multiplier for a more structured uncertainty. We remark that it causes no loss of generality to assume $0\in\DeltaB_l$ for all $l\in\Z$.

Now we are able to state and prove numerically tractable conditions which ensure that the conditions of the robust counterparts of Lemma~\ref{theo:as} and Theorem~\ref{theo:diss2P} are satisfied and therefore imply robust asymptotic stability and performance. We start with robust asymptotic stability.

\begin{theorem}
\label{thm:robStab}
The LFR of the CSLS is well-posed and there exist matrices $X_i\succ 0$ for all $i\in\V$ such that\alnl{\label{eq:RobStabLyap}\tilde{A}_l({\Delta}_l)\tp X_j \tilde{A}_l({\Delta}_l)-X_i\prec 0} for all $\Delta_l\in\DeltaB_l$ and all $(i,j,l)\in\E$ if there exist $X_i\succ 0$ for all $i\in\V$ such that\alnl{\label{eq:robStabdiss}(\bullet)\tp\mat{cc}{-X_i&0\\0&X_j}\mat{cc}{I&0\\A_l&B^{w_u}_{l}}+(\bullet)\tp P_l^{\Delta}\mat{cc}{0&I\\C^{z_u}_{l}&D^{z_uw_u}_{l}}\prec0}for a $P_l^{\Delta}\in\P_l^{\Delta}$ and for all $(i,j,l)\in\E$.  

In the case that $\DeltaB_l$ is compact for all $l\in\Z$, also the converse holds.
\end{theorem}
\proofen{The proof follows standard arguments. Fix some $(i,j,l)\in\E$. The right lower block of~\eqref{eq:robStabdiss} implies\aln{\mat{c}{I\\D^{z_uw_{u}}_{l}}\tp P_l^{\Delta} \mat{c}{I\\D^{z_uw_{u}}_{l}}\prec 0.}Together with~\eqref{eq:SepUn} this implies that $I-D^{z_uw_{u}}_{l}\Delta_l$ is invertible for $\Delta_l\in\DeltaB_l$. Right- and left multiplying $\mathrm{row}(I,(\Delta_l(I-D^{z_uw_{u}}_{l}\Delta_l)\inv C_l))$ on~\eqref{eq:robStabdiss} yields\aln{
0&\succ(\bullet)\tp\mat{cc}{-X_i&0\\0&X_j}\mat{c}{I\\\tilde{A}_l({\Delta_l})}+(\bullet)\tp P_l^{\Delta}\mat{c}{\Delta_l\\I}(I-D^{z_uw_{u}}_{l}\Delta_l)\inv C_l\\&\succeq \tilde{A}_l({\Delta}_l)\tp X_j \tilde{A}_l({\Delta}_l)-X_i}for all $\Delta_l\in\DeltaB_l$. Because this holds for all $(i,j,l)\in\E$ the claim follows.

The proof for the converse statement in the case of $\DeltaB_l$ compact for all $l\in\Z$ is an application of the full block S-procedure. It only requires minor modifications of the proof in the case of linear systems~\cite{Scherer2000} and arguments which are routine after our discussion. For the reader's convenience, we provide a sketch of the proof. Fix some $(i,j,l)\in\E$. Now define the matrices\aln{
N&=\mat{cc|cc}{-X_i&0&0&0\\0&X_j&0&0\\\hline 0&0&0&0\\0&0&0&0}, \quad
S=\im\mat{cc}{I&0\\A_l&B_l\\\hline0&I\\C_l&D_l}, \quad
S_0=\im\mat{c}{0\\B_l\\\hline I\\D_l},\\
U&=\mat{cc}{I&-\Delta_i}, \quad
T=\mat{cc|cc}{0&0&I&0\\0&0&0&I}.}A calculation shows that well-posedness and~\eqref{eq:RobStabLyap} are equivalent to $N\prec0$ on the subset $S\cap\ker(UT)$ and\lbreak$S_0\cap S\cap\ker(UT)=\{0\}$. Applying the full block S-procedure gives the alternative conditions $N+T\tp P_l^{\Delta} T \prec 0$ on $S$ and $P_l^{\Delta} \succ 0$ on $\ker(U)$. The former alternative condition is nothing but~\eqref{eq:robStabdiss} and the latter is~\eqref{eq:SepUn}. Because this holds for all $(i,j,k)\in\E$ the claim follows.}

Observe that Theorem~\ref{thm:robStab} reformulates the problem of finding solutions of coupled Lyapunov inequalities for uncertain system representations into a dissipativity property for a given state-space representation. For a fixed multiplier index $(P_l^\Delta)_{l=1}^m$, the index can be considered as a dissipativity index and hence the reformulations provided by Theorem~\ref{thm:main} and Corollary~\ref{thm:mainC} can be used. Moreover, revising the steps of the corresponding proofs shows that it causes no harm to vary the dissipativity index in a given set. In other words, we can use the formulations in Theorem~\ref{thm:main} and Corollary~\ref{thm:mainC}, under the assumption that the requirements are satisfied for each $P_l^{\Delta}\in\P_l^{\Delta}$ for all $l\in\Z$, to search for certificates $X_i, \tilde{X}_i, G_i$, $i\in\V$ and multipliers $P_l^{\Delta}$, $l\in\Z$ simultaneously.

We have seen that the problem of guaranteeing quadratic performance or robust stability can be viewed as special instances of the more general problem of certifying dissipativity of a CSLS. This motivates to reformulate the problem of guaranteeing robust quadratic performance of a CSLS into the problem of verifying a dissipativity property. We show this reformulation in the next Theorem.
\begin{theorem}
\label{thm:robPer}Given a dissipativity index $P$ with $R_l\succeq0,~l\in\Z$. If there exist $X_i\succ 0$ for all $i\in\V$ such that\alnl{
\begin{split}\label{eq:robPerdiss}(\bullet)\tp\mat{ccc}{-X_i&0\\0&X_j}\mat{ccc}{I&0&0\\A_l&B^{w_u}_{l}&B^{w_p}_{l}}&+(\bullet)\tp P_l^{\Delta}\mat{ccc}{0&I&0\\C^{z_u}_{l}&D^{z_uw_u}_{l}&D^{z_uw_p}_{l}}\\&+(\bullet)\tp P_l\mat{ccc}{0&0&I\\C^{z_p}_{l}&D^{z_pw_u}_{l}&D^{z_pw_p}_{l}}\prec0\end{split}} for a $P_l^{\Delta}\in\P_l^{\Delta}$ and all $(i,j,l)\in\E$, then the LFR of the CSLS is well-posed and there exist matrices $X_i\succ 0$ for all $i\in\V$ such that~\eqref{eq:RobStabLyap} holds and the matrices certify robust dissipativity with index $(P_l)_{l=1}^m$.

In the case that $\DeltaB_l$ is compact for all $l\in\Z$, also the converse holds.
\end{theorem}

\proofen{Inequality~\eqref{eq:robPerdiss} implies validity of~\eqref{eq:robStabdiss}. Together with Theorem~\ref{thm:robStab} we infer that~\eqref{eq:RobStabLyap} holds and that the CSLS is well-posed. It is obvious that~\eqref{eq:robPerdiss} stays true if we uniformly replace $P_l$ for all $l\in\V$ with $P_l+\mat{cc}{\epsilon I&0\\0&0}$ for a sufficiently small $\epsilon>0$. For a fixed $(i,j,l)\in\E$ and $t\geq0$ we can multiply the perturbed inequality with an admissible trajectory $\mathrm{col}(x(t), w_u(t), w_p(t))$ and obtain\aln{
x(t+1)\tp X_j x(t+1) - x(t)\tp X_ix(t)&+z_u(t)\tp\mat{c}{\Delta_l\tp\\I}\tp P_l^{\Delta}\mat{c}{\Delta_l\\I}z_u(t)\\&+\mat{c}{w_p(t)\tp\\z_p(t)\tp}\tp P_l\mat{c}{w_p(t)\\z_p(t)}\leq -\epsilon w_p(t)\tp w_p(t).}We exploited that $w_u(t)=\Delta_lz_u(t)$ for $\Delta_l\in\DeltaB_l$. Together with~\eqref{eq:SepUn}, we obtain \aln{
x(t+1)\tp X_j x(t+1) - x(t)\tp X_i x(t)+\mat{c}{w_p(t)\tp\\z_p(t)\tp}\tp P_l\mat{c}{w_p(t)\\z_p(t)}\leq -\epsilon w_p(t)\tp w_p(t).}Because we considered an arbitrary admissible trajectory at an arbitrary $t\geq0$ for all $(i,j,l)\in\E$ the claim follows.

Similar to the proof of Theorem~\ref{thm:robStab}, the proof of the converse follows the same arguments, mutatis mutandis, as for linear systems and is omitted here. The proof in the case of linear systems can be found, e.g., in~\cite{Scherer2000}.}

Note, that Theorem~\ref{thm:robPer} together with a robust version of Theorem~\ref{theo:diss2P} implies robust quadratic performance of the CSLS. The conditions in Theorem~\ref{thm:robPer} can be interpreted as a dissipativity condition on a nominal CSLS. If we define a CSLS $\bar{\Sigma}$ with the same graph and a family of state-space representations which assign to $l\in\Z$ the state-space representation\aln{(A_l,\mathrm{col}(B^{w_u}_{l},B^{w_p}_{l}),\mathrm{row}(C^{z_u}_{l},C^{z_p}_{l}),\mathrm{row}(\mathrm{col}(D^{z_uw_u}_{l}, D^{z_uw_p}_{l}),\mathrm{col}(D^{z_pw_u}_{l},D^{z_pw_p}_{l}))),}then the conditions in Theorem~\ref{thm:robPer} can be interpreted as the requirement of asymptotic stability and dissipativity with index\aln{(\bar{P}_l)_{l=1}^m=\left(\mat{cc|cc}{Q_l^{\Delta}&0&S_l^{\Delta}&0\\0&Q_l&0&S_l\\\hline(S_l^{\Delta})\tp&0&R_l^{\Delta}&0\\0&S_l\tp&0&R_l}\right )_{l=1}^m} for $\bar{\Sigma}$. In the last equation, we partition the matrix $P_l^{\Delta}=\mat{cc}{Q_l^{\Delta}&S_l^{\Delta}\\(S_l^{\Delta})\tp&R_l^{\Delta}}$ for each $l\in\V$ according to the structure induced by~\eqref{eq:SepUn}. In this way, the reformulation provided by Theorem~\ref{thm:main} and Corollary~\ref{thm:mainC} can be used to obtain numerically tractable conditions for robust quadratic performance.

There might be the question of why we use a trajectory based argument and the full block S-procedure in the proofs of Theorem~\ref{thm:robStab} and~\ref{thm:robPer} instead of using purely the full block S-procedure. The reason for that is that in this way, everything in this section except for the statements about necessity in Theorem~\ref{thm:robStab} and~\ref{thm:robPer} remains true if the uncertainty is a nonlinear and possible time-varying operator. The only thing which needs attention is that Theorem~\ref{thm:robStab} and~\ref{thm:robPer} do not imply well-posedness of the LFR of the CSLS anymore and this has to be guaranteed additionally, e.g., with standard arguments of ordinary differential equation theory. In the case that $D^{z_uw_u}_{l}=0$ for all $l\in\Z$, well-posedness of the LFR of the CSLS is immediate.
\newpage\section{SYNTHESIS}\label{sec:synthesis}
In the last section, we considered the problem of analyzing a CSLS with respect to (robust) quadratic performance using LMIs. Moreover, we have seen that these problems can be reduced to verifying a dissipativity property. In this section, we aim to answer the question of how we can use convex optimization to construct controllers which control a CSLS in such a way that it is asymptotically stable and dissipative with a given index. The latter property can then be interpreted, e.g., as a robustness or performance property of the controlled CSLS. 

For each label $l$ of a given constraining graph, we consider the state-space representation of the uncontrolled system which is assigned to $l$ given by\aln{x(t+1)&=A^{u}_{l}x(t)+B_lw(t)+B^{u}_{l}u(t)\\y(t)&=C^{u}_{l}x(t)+D_lw(t)+D^{yu}_{l}u(t)} where $u(t)$ denotes the control input at time $t$. The CSLS which is defined by such a family of state-space representations and a given labeled graph will be called ``uncontrolled CSLS'' in the sequel. We want to design a node-dependent controller $u(t)=K_ix(t)$ such that the controlled CSLS is asymptotically stable and dissipative with a predefined index. For each $i\in\V$ and $l\in\Z$ such that there exists a $j\in\V$ with $(i,j,l)\in\E$, we can close the loop and obtain the representation\aln{x(t+1)&=A_{i,l}x(t)+B_lw(t)=(A^{u}_l+B^{u}_lK_i)x(t)+B_lw(t)\\y(t)&=C_{i,l}x(t)+D_lw(t)=(C^{u}_l+D^{yu}_lK_i)x(t)+D_lw(t).}In contrast to Definition~\ref{def:CSLS}, the system matrices do not only depend on the current label but also on the node. However, this causes no harm to the results in the previous two sections as an inspection reveals. We know that a CSLS is asymptotically stable and dissipative with a given index if one of the conditions in Theorem~\ref{thm:main} or Corollary~\ref{thm:mainC} is satisfied. If we plug the expression of the closed-loop in these conditions, then the resulting inequalities are not affine in the node-dependent controller-gain and the certificate simultaneously. However, these nonlinearities can be easily removed by a variable substitution. This is stated in the next theorem and corollary.
\begin{theorem}
\label{thm:mainSyn}Given an uncontrolled CSLS and an index $(P_l)_{l=1}^{m}$ with $R_l\succeq0$, $l\in\Z$ and the decomposition $R_l=U_l\tp\tilde{R}\inv U_l$ with $\tilde{R}\inv_l\succ0$, then there exist $X_i\succ 0$ and a node-dependent state-feedback gain $K_i$ for all $i\in\V$ such that\aln{A_{i,l}\tp X_j A_{i,l}-X_i\prec 0} for all $(i,j,l)\in\E$ and the controlled CSLS is dissipative with index $(P_l)_{l=1}^{m}$ if one of the following equivalent statements is satisfied:\newpage
\begin{enumerate}[(i)]
	\item \label{eq:char3Syn} There exist $\tilde{X}_i=\tilde{X}_i\tp$ and $Z_i$ for all $i\in\V$ such that\alnl{\label{eq:mainSSyn}\mat{cccc}{\tilde{X}_j&A^{u}_l\tilde{X}_i+B^{u}_lZ_i&B_l&0\\
																																					 \bullet&\tilde{X}_i&-(C^u_l\tilde{X}_i+D^{yu}_lZ_i)\tp S_l\tp&(C^u_l\tilde{X}_i+D^{yu}_lZ_i)\tp U_l\tp\\
																																					 \bullet&\bullet&-Q_l-S_lD_l-D_l\tp S_l\tp&D_l\tp U_l\tp\\
																																				   \bullet&\bullet&\bullet&\tilde{R}_l}\succ 0}for all $(i,j,l)\in\E$.																																				
	\item \label{eq:char4Syn} There exist $\tilde{X}_i=\tilde{X}_i\tp$, $G_i$, and ${Z}_i$ for all $i\in\V$ such that\alnl{\label{eq:mainPSyn}\mat{cccc}{\tilde{X}_j&A^{u}_lG_i+B^u_l Z_i&B_l&0\\
																																									 \bullet&G_i+G_i\tp-\tilde{X}_i&-(C^{u}_lG_i+D^{yu}_l Z_i)\tp S_l\tp&(C^{u}_lG_i+D^{yu}_l Z_i)\tp U_l\tp\\
																																									 \bullet&\bullet&-Q_l-S_lD_l-D_l\tp S_l\tp&D_l\tp U_l\tp\\
																																									 \bullet&\bullet&\bullet&\tilde{R}_l}\succ 0}for all $(i,j,l)\in\E$.
\end{enumerate}The controller is given by $K_i=Z_i\tilde{X}_i\inv$ or $K_i={Z}_iG_i\inv$, respectively.
\end{theorem}
\proofen{The statement follows directly by plugging the state-space representations of the closed-loop into the conditions of Theorem~\ref{thm:main} and defining $Z_i=K_iX_i$ in Condition~\eqref{eq:char3Syn} and ${Z}_i=K_iG_i$ in Condition~\eqref{eq:char4Syn} for all $i\in\V$. The formulas for the controllers follow from the definition and the fact that~\eqref{eq:mainSSyn} implies that $X_i$ is invertible for all $i\in\V$ and~\eqref{eq:mainPSyn} implies the same for $G_i$.}
\begin{corollary}
\label{thm:mainCSyn}Under the assumptions of Theorem~\ref{thm:mainSyn} and the additional assumption that every element of $(P_l)_{l=1}^{m}$ is invertible, the equivalent conditions~\eqref{eq:char3Syn}-\eqref{eq:char4Syn} of Theorem~\ref{thm:mainSyn} are equivalent to the following equivalent conditions:
\begin{enumerate}[(i)]
	\item \label{eq:char2DSyn} There exist $\tilde{X}_i=\tilde{X}_i\tp$ and $Z_i$ for all $i\in\V$ such that\alnl{\label{eq:mainSDSyn}
				\mat{ccc}{\tilde{X}_i&(A^{u}_l\tilde{X}_i+B^{u}_lZ_i)\tp&(C^{u}_l\tilde{X}_i+D^{yu}_lZ_i)\tp\\
				         \bullet&\tilde{X}_j+B_l\tilde{Q}B_l\tp&-B_l\tilde{S}_l+B_l\tilde{Q}_lD_l\tp\\
								 \bullet&\bullet&\tilde{R}_l-\tilde{S}_lD_l-D_l\tp\tilde{S}_l\tp+D_l\tilde{Q}_lD_l\tp}\succ 0}for all $(i,j,l)\in\E$ and $\tilde{Q_l}\preceq0$ for all $l\in\Z$.
	\item \label{eq:char3DSyn} There exist $\tilde{X}_i=\tilde{X}_i\tp$, $G_i$, and $Z_i$ for all $i\in\V$ such that\alnl{\label{eq:mainPDSyn}
				\mat{ccc}{G_i+G_i\tp-\tilde{X}_i&(A^{u}_lG_i+B^{u}_lZ_i)\tp&(C^{u}_lG_i+D^{yu}_lZ_i)\tp\\
				         \bullet&\tilde{X}_j+B_l\tilde{Q}B_l\tp&-B_l\tilde{S}_l+B_l\tilde{Q}_lD_l\tp\\
								 \bullet&\bullet&\tilde{R}_l-\tilde{S}_lD_l-D_l\tp\tilde{S}_l\tp+D_l\tilde{Q}_lD_l\tp}\succ 0}for all $(i,j,l)\in\E$ and $\tilde{Q_l}\preceq0$ for all $l\in\Z$.	
			
\end{enumerate}The controller is given by $K_i=Z_iX_i\inv$ or $K_i={Z}_iG_i\inv$, respectively.
\end{corollary}
\proofen{The proof follows along the same lines as the proof of Theorem~\ref{thm:mainSyn} but applies the arguments on the conditions of Corollary~\ref{thm:mainC}.} Now we are able to design a node-dependent controller such that a set of coupled Lyapunov inequalities is satisfied and the controlled CSLS is dissipative with a given index. In view of our discussion in Section~\ref{sec:analysis}, this means that we are able to find a node-dependent controller such that the controlled CSLS satisfies a given performance criterion either nominally or robustly. In the case that the controlled CSLS should satisfy a performance criterion nominally, there is no reason to prefer either Theorem~\ref{thm:mainSyn} or Corollary~\ref{thm:mainCSyn} for synthesis. However, if we are interested in robust stability or performance, Subsection~\ref{sec:robustness} has revealed that it is beneficial to optimize over the multiplier index which can be interpreted as a dissipativity index. Therefore, the conditions of Corollary~\ref{thm:mainCSyn} are to be preferred because they allow such a simultaneous optimization. This requires that all considered multipliers are invertible and a convex description of the set of all inverses is available. Clearly, it is also possible to consider a convex subset of the set of the inverses of the considered multipliers but in general, this will introduce conservatism. A possible way to obtain such a convex description is, in the case that the assumptions are satisfied, an application of the Dualization Lemma to the description of the multiplier set. At this point, we want to highlight that such a convex description is also needed for multiplier based robust state-feedback design for linear systems.

It might be the case that in some situations a node-independent controller should be designed. This can be easily achieved by setting $G_i=G$ and $Z_i=Z$ for all $i\in\V$ in Condition~\eqref{eq:char4Syn} of Theorem~\ref{thm:mainSyn} or Corollary~\ref{thm:mainCSyn}. Clearly, this approach, which is well-known in the literature, e.g.~\cite{Linsenmayer2017}, introduces conservatism. However, the introduced conservatism is smaller compared to the case where we set $X_i=X$ and $Z_i=Z$ for all $i\in\V$ in Condition~\eqref{eq:char2DSyn} of Theorem~\ref{thm:mainSyn} or Corollary~\ref{thm:mainCSyn} because the latter means that we are looking for a global quadratic node-independent storage function.

So far we have only discussed how to design controllers which achieve (robust) quadratic performance. In Section~\ref{sec:analysis} we used the generalized $\mathcal{H}^2$-norm to show how the presented framework allows us to deal with performance criteria which are beyond quadratic performance. This also holds for synthesis. It is not hard to see that the convexifying substitutions of Theorem~\ref{thm:mainSyn} can be used to derive a convex optimization problem for designing a controller which renders the controlled CSLS such that it satisfies a generalized $\mathcal{H}^2$-performance criterion. Moreover, we can perform similar arguments as in Section~\ref{sec:robustness} to obtain analysis inequalities for robust generalized $\mathcal{H}^2$-performance which can be transformed into a convex synthesis problem with the same arguments.
\section{APPLICATION TO WEAKLY HARD REAL-TIME CONTROL SYSTEMS}\label{sec:WHRT}
In this section, we apply the developed methods to the problem of designing state-feedback controllers for WHRT control systems.
In those kinds of systems, the losses in the feedback loop described by a WHRT constraint affect the stability and control performance.
As motivated in the introduction, a WHRT control system can, for example, represent an NCS with packet dropouts due to network effects, or a control task in a real-time control setting using a shared computation resource.
Using WHRT control systems is already well established in the literature, see the survey~\cite{Salamun2023} for an overview,~\cite{Blind2015,Linsenmayer2017,Linsenmayer2021,Jia2005} for WHRT constraints in NCS, and~\cite{Ahrendts2018,Maggio2020,Xu2023} for WHRT constraints in real-time control settings.

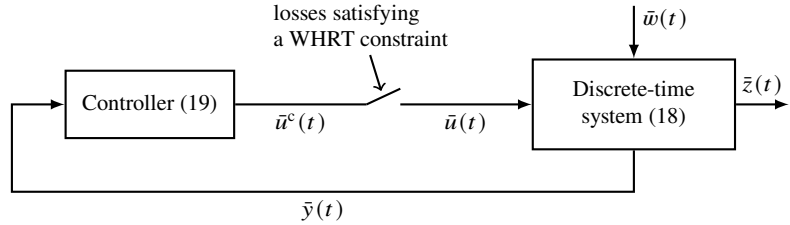
\begin{figure}
	\centering
	\begin{tikzpicture}[auto,>=latex,thick]
		\node [block,minimum width=9em,minimum height=4em,align=center] (sys) {Discrete-time\\system \eqref{eq:WHRTsystem}};
		\draw [-latex] (sys.north) +(0,0.7) -- node[right, pos=0.25] {$\bar{w}(t)$} (sys.north);
		\draw [-latex] (sys.east) -- node[above, pos=0.5] {$\bar{z}(t)$} +(0.7,0);
		\coordinate [left=of sys,xshift=-2.5em](dropRight);
		\coordinate (dropLeft) at ($(dropRight)+(-1.5em,0)$);
		\node [block,left=of dropLeft,minimum width=7.5em,xshift=-2.5em] (K) {Controller \eqref{eq:WHRTcon}};
		\draw [-latex] (sys.south) |- node[below, pos=0.75] {$\bar{y}(t)$} ($(K.south west)+(-0.7,-0.7)$) |- (K.west);
		\draw (K.east) --  node[below, pos=0.5] {{$\bar{u}^\mathrm{c}(t)$}} (dropLeft);
		\draw (dropLeft) -- ($(dropRight)+(0,0.7em)$);
		\draw [-latex] (dropRight) --  node[below, pos=0.5] {$\bar{u}(t)$} (sys.west);
		\node[above,align=left] (desc) at ($(dropLeft)!0.5!(dropRight)+(-1em,2.2em)$) {losses satisfying\\a WHRT constraint};
		\draw[-{Classical TikZ Rightarrow}] (desc) -- ($(dropLeft)!0.5!(dropRight)+(-0.1em,0.6em)$);
	\end{tikzpicture}
	\label{fig:blockschaltbild-WHRTsystem}
	\caption{WHRT control system with losses in the feedback loop.}
\end{figure}

In the following, we show, exemplary, how a problem in networked control can be reformulated into a WHRT control system and how the presented theory can be applied to it. The same approach can also be evoked for a similar real-time control setting. We consider the discrete-time linear time-invariant system represented by\alnl{
\nonumber\bar{x}(t+1)&=\bar{A}\bar{x}(t)+\bar{B}\bar{w}(t)+\bar{B}^{\bar{u}}\bar{u}(t)\\
\label{eq:WHRTsystem}\bar{y}(t)&=\bar{x}(t)\\
\nonumber\bar{z}(t)&=C\bar{x}(t)+\bar{D}\bar{w}(t)+\bar{D}^{\bar{u}}\bar{u}(t)}with initial condition $\bar{x}(0)=\bar{x}_0$, control input $\bar{u}$, performance input $\bar{w}$, measurement output $\bar{y}$, and performance output $\bar{z}$. The goal is to find a controller\alnl{\label{eq:WHRTcon}\bar{u}^c(t)=K\bar{y}(t)} such that the resulting closed-loop system, i.e.,~the system which results from setting $\bar{u}=\bar{u}^c$, is asymptotically stable and satisfies a desired quadratic performance criterion specified by a symmetric matrix $P$, i.e., \aln{\sum \limits_{t=0}^{\infty}\mat{c}{\bar{w}(t)\\\bar{z}(t)}\tp P\mat{c}{\bar{w}(t)\\\bar{z}(t)}\leq-\tilde{\epsilon}\norm{w}_{\ell^2}} for $\bar{x}_0=0$ and all $w\in \ell^2$. Note that this performance specification includes also the classical criterion on the $\ell^2$-gain or passivity. In addition to the described standard setup, and in view of NCS, we assume that the control input $u^c(t)$ is not applied to the system at every time instant, but may get lost occasionally.
This can be the case if, for example, the system and controller are connected via unreliable communication channels where transmitted packets can get lost c.f.~Figure~\ref{fig:blockschaltbild-WHRTsystem}.
There are two main strategies in the literature on how the actuator receiving the control signal can handle such losses of control information. The first one sets $\bar{u}(t)=0$ at such time instants and is called zero strategy.
The other strategy uses the last received control input at such time instants and is referred to as the hold strategy. Already for simple scalar systems there exist examples which show that none of the two is superior to the other~\cite{Schenato2009}.

If the uncontrolled system is not asymptotically stable or does not satisfy the desired performance criterion and we do not specify the losses in more detail, then there exists no controller which renders the closed-loop asymptotically stable and satisfies the desired quadratic performance criterion.
However, from an application oriented perspective, it makes sense to specify the losses in more detail and to take this description into account for controller synthesis.
In the literature, there exist different approaches to model such losses, e.g., with Bernoulli distributions and their generalization to Markov chains or deterministic descriptions such as WHRT constraints. Describing losses by deterministic descriptions possesses the advantages that deterministic guarantees on the performance of the closed-loop can be given. Such deterministic guarantees are crucial, e.g., for safety-critical control tasks.
Assuming that a loss sequence satisfies a WHRT constraint may seem a strong assumption, however there exist multiple techniques for designing and scheduling the underlying network, such that the satisfaction of such a constraint can easily be guaranteed, see e.g., \cite{Ahrendts2018, Broster2002, Hamdaoui1995}.
There are different types of WHRT constraints~\cite{Bernat2001}. In their essence, they describe how many control attempts are at least successful within a moving time window, i.e., how often $\bar{u}(t)=\bar{u}^c(t)$ within the time window. In this way, a WHRT constraint also provides an upper bound of consecutive unsuccessful control attempts (time steps $t\in\N$ for which $\bar{u}(t)\neq\bar{u}^c(t)$) which we will denote by $m-1$. The reason why we denote it by $m-1$ and not by $m$ will become clearer later in this paragraph. The system defined by the state-space representation~\eqref{eq:WHRTsystem}, the controller~$\eqref{eq:WHRTcon}$, and the set of all possible sequences of losses described by the WHRT constraint is called WHRT control system.

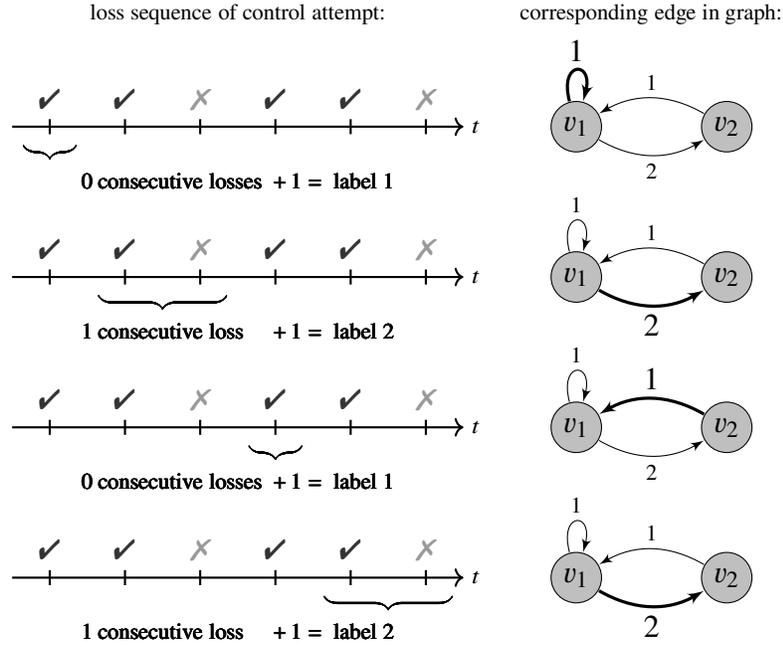
\begin{figure}
	\centering
	\begin{tikzpicture}
		\colorlet{darkGreen}{white!20!black}
		\colorlet{redCross}{white!60!black}
		\tikzset{vertexFill/.style = {shape=circle,fill=gray!50!white,draw,minimum size=1.5em,inner sep=0.3em},thick}
		\tikzset{edge/.style = {->,> = latex'}}
		\tikzset{edgeMarked/.style = {->,> = latex',very thick}}
		
		\def\numAttempt{6}
		\def\spacingAxis{1.0}
		\def\sideLengthBraces{0.35}
		
		\node[] at (2.5,1.5) {loss sequence of control attempt:};
		\node[] at (8,1.5) {corresponding edge in graph:};
		
		\begin{scope}[shift={(0,0)}]
			\draw[thick, ->] (-0.5*\spacingAxis,0) -- (\numAttempt*\spacingAxis-0.5*\spacingAxis,0);
			\node[right] at (\spacingAxis*\numAttempt-0.5*\spacingAxis,0) {$t$};
			\foreach \k / \label in {0/1, 1/1, 2/0, 3/1, 4/1, 5/0} {
				\ifnum \label=0
					\node[above,color=redCross] at (\k*\spacingAxis,0.1) {\XSolidBrush};
				\else
					\node[above,color=darkGreen] at (\k*\spacingAxis,0.1) {\CheckmarkBold};
				\fi
				
				\draw[thick] (\k*\spacingAxis,0.1) -- (\k*\spacingAxis,-0.1);
				\node[below] at (\k*\spacingAxis,-0.1) (attempt\k) {};
				
				\draw[decorate,thick,decoration={calligraphic brace,amplitude=5pt,mirror}] (0*\spacingAxis-\sideLengthBraces*\spacingAxis,-0.25) -- (0*\spacingAxis+\sideLengthBraces*\spacingAxis,-0.25);
				\node[below] at (0.5*\numAttempt*\spacingAxis-0.5*\spacingAxis,-0.5) {$0 \text{ consecutive losses } + 1 = \text{ label } 1$};
			}
			
			\begin{scope}[shift={(\numAttempt*\spacingAxis+\spacingAxis,0)}]
				\node[vertexFill] (v1) at (0,0) {\large$v_1$};
				\node[vertexFill] (v2) at (2,0) {\large$v_2$};
				
				\draw[edgeMarked] (v1) to[loop above] node[above, pos=0.5] (v1label) {\large$1$} (v1);
				\draw[edge] (v1) to[bend right] node[below, pos=0.5] (v2label) {$2$} (v2);
				\draw[edge] (v2) to[bend right] node[above, pos=0.5] (v3label) {$1$} (v1);
			\end{scope}
		\end{scope}
		
		\begin{scope}[shift={(0,-2)}]
			\draw[thick, ->] (-0.5*\spacingAxis,0) -- (\numAttempt*\spacingAxis-0.5*\spacingAxis,0);
			\node[right] at (\spacingAxis*\numAttempt-0.5*\spacingAxis,0) {$t$};
			\foreach \k / \label in {0/1, 1/1, 2/0, 3/1, 4/1, 5/0} {
				\ifnum \label=0
					\node[above,color=redCross] at (\k*\spacingAxis,0.1) {\XSolidBrush};
				\else
					\node[above,color=darkGreen] at (\k*\spacingAxis,0.1) {\CheckmarkBold};
				\fi
				
				\draw[thick] (\k*\spacingAxis,0.1) -- (\k*\spacingAxis,-0.1);
				\node[below] at (\k*\spacingAxis,-0.1) (attempt\k) {};
				
				\draw[decorate,thick,decoration={calligraphic brace,amplitude=5pt,mirror}] (1*\spacingAxis-\sideLengthBraces*\spacingAxis,-0.25) -- (2*\spacingAxis+\sideLengthBraces*\spacingAxis,-0.25);
				\node[below] at (0.5*\numAttempt*\spacingAxis-0.5*\spacingAxis,-0.5) {$1 \text{ consecutive loss\phantom{es} } + 1 = \text{ label } 2$};
			}
			
			\begin{scope}[shift={(\numAttempt*\spacingAxis+\spacingAxis,0)}]
				\node[vertexFill] (v1) at (0,0) {\large$v_1$};
				\node[vertexFill] (v2) at (2,0) {\large$v_2$};
				
				\draw[edge] (v1) to[loop above] node[above, pos=0.5] (v1label) {$1$} (v1);
				\draw[edgeMarked] (v1) to[bend right] node[below, pos=0.5] (v2label) {\large$2$} (v2);
				\draw[edge] (v2) to[bend right] node[above, pos=0.5] (v3label) {$1$} (v1);
			\end{scope}
		\end{scope}
		
		\begin{scope}[shift={(0,-4)}]
			\draw[thick, ->] (-0.5*\spacingAxis,0) -- (\numAttempt*\spacingAxis-0.5*\spacingAxis,0);
			\node[right] at (\spacingAxis*\numAttempt-0.5*\spacingAxis,0) {$t$};
			\foreach \k / \label in {0/1, 1/1, 2/0, 3/1, 4/1, 5/0} {
				\ifnum \label=0
					\node[above,color=redCross] at (\k*\spacingAxis,0.1) {\XSolidBrush};
				\else
					\node[above,color=darkGreen] at (\k*\spacingAxis,0.1) {\CheckmarkBold};
				\fi
				
				\draw[thick] (\k*\spacingAxis,0.1) -- (\k*\spacingAxis,-0.1);
				\node[below] at (\k*\spacingAxis,-0.1) (attempt\k) {};
				
				\draw[decorate,thick,decoration={calligraphic brace,amplitude=5pt,mirror}] (3*\spacingAxis-\sideLengthBraces*\spacingAxis,-0.25) -- (3*\spacingAxis+\sideLengthBraces*\spacingAxis,-0.25);
				\node[below] at (0.5*\numAttempt*\spacingAxis-0.5*\spacingAxis,-0.5) {$0 \text{ consecutive losses } + 1 = \text{ label } 1$};
			}
			
			\begin{scope}[shift={(\numAttempt*\spacingAxis+\spacingAxis,0)}]
				\node[vertexFill] (v1) at (0,0) {\large$v_1$};
				\node[vertexFill] (v2) at (2,0) {\large$v_2$};
				
				\draw[edge] (v1) to[loop above] node[above, pos=0.5] (v1label) {$1$} (v1);
				\draw[edge] (v1) to[bend right] node[below, pos=0.5] (v2label) {$2$} (v2);
				\draw[edgeMarked] (v2) to[bend right] node[above, pos=0.5] (v3label) {\large$1$} (v1);
			\end{scope}
		\end{scope}
		
		\begin{scope}[shift={(0,-6)}]
			\draw[thick, ->] (-0.5*\spacingAxis,0) -- (\numAttempt*\spacingAxis-0.5*\spacingAxis,0);
			\node[right] at (\spacingAxis*\numAttempt-0.5*\spacingAxis,0) {$t$};
			\foreach \k / \label in {0/1, 1/1, 2/0, 3/1, 4/1, 5/0} {
				\ifnum \label=0
					\node[above,color=redCross] at (\k*\spacingAxis,0.1) {\XSolidBrush};
				\else
					\node[above,color=darkGreen] at (\k*\spacingAxis,0.1) {\CheckmarkBold};
				\fi
				
				\draw[thick] (\k*\spacingAxis,0.1) -- (\k*\spacingAxis,-0.1);
				\node[below] at (\k*\spacingAxis,-0.1) (attempt\k) {};
				
				\draw[decorate,thick,decoration={calligraphic brace,amplitude=5pt,mirror}] (4*\spacingAxis-\sideLengthBraces*\spacingAxis,-0.25) -- (5*\spacingAxis+\sideLengthBraces*\spacingAxis,-0.25);
				\node[below] at (0.5*\numAttempt*\spacingAxis-0.5*\spacingAxis,-0.5) {$1 \text{ consecutive loss\phantom{es} } + 1 = \text{ label } 2$};
			}
			
			\begin{scope}[shift={(\numAttempt*\spacingAxis+\spacingAxis,0)}]
				\node[vertexFill] (v1) at (0,0) {\large$v_1$};
				\node[vertexFill] (v2) at (2,0) {\large$v_2$};
				
				\draw[edge] (v1) to[loop above] node[above, pos=0.5] (v1label) {$1$} (v1);
				\draw[edgeMarked] (v1) to[bend right] node[below, pos=0.5] (v2label) {\large$2$} (v2);
				\draw[edge] (v2) to[bend right] node[above, pos=0.5] (v3label) {$1$} (v1);
			\end{scope}
		\end{scope}
	\end{tikzpicture}%
	\caption{Construction of the underlying graph of the CSLS for a WHRT control system with the WHRT constraint ``two successful control attempts within every window of length 3''.}
	\label{fig:graph-construction}
\end{figure}

The idea is now to consider the described setup as a CSLS where the labels correspond to the number of consecutive unsuccessful control attempts between two successful ones plus one. The ``plus one'' takes into account that between two sequences of consecutive unsuccessful control attempts there has to be one successful control attempt. Otherwise, both sequences can be concatenated. In this way, $m$ labels are defined. This also clarifies why we denoted the upper bound of consecutive unsuccessful control attempts by $m-1$. An example for the construction of such a graph is given in Figure~\ref{fig:graph-construction}.
If we assign to each label a system representation which describes the behavior of the WHRT control system when the specified number of consecutive losses occurs, then the WHRT control system is reformulated into a CSLS. The described approach was proposed for stabilization in~\cite{Linsenmayer2017} and there is an algorithmic method available which reformulates a WHRT constraint into the corresponding constraining graph~\cite{Linsenmayer2017, Linsenmayer2021}. We are not only interested in stability properties of the closed-loop but also in performance properties. However, the mentioned approach does not directly apply to systems with performance channels because between two successful consecutive control attempts, there can be a varying number, up to $m-1$, of time steps where no control attempt is successful. At these time steps the system is still influenced by the performance input and also generates a performance output value. Therefore, we collect all input and output values between two consecutive successful control attempts in two vectors and consider them as an input and output value of a CSLS. Recall that the constraining graph is already defined by the WHRT constraint. We used the same idea for nominal $\ell^2$-performance of weakly hard real-time control systems in~\cite{Seidel2023}. This approach results in a CSLS where the systems linked to the different labels have different input and output dimensions. Note that if $\bar{w}$ and $\bar{z}$ denote the input and output trajectory of the WHRT control system and $e$ the switching sequence of the corresponding CSLS, then the input and output trajectory of the CSLS is given by $w=q_{d_\mathrm{i}(\sigma(e))}(\bar{w})$ and $z=q_{d_\mathrm{o}(\sigma(e))}(\bar{z})$ where $q$ denotes the isomorphism of Section~\ref{sec:subnominal}. Deriving state-space representations of the systems which are linked to the different labels is an exploitation of linearity of~\eqref{eq:WHRTsystem}. For the zero strategy we obtain the representation\alnl{\nonumber
x(t+1)&=\bar{A}^{l}x(t)+\mat{cccc}{\bar{A}^{l-1}\bar{B}&\bar{A}^{l-2}\bar{B}&\hdots&\bar{B}}w(t)+\bar{A}^{l-1}\bar{B}^{\bar{u}}u(t)\\\label{eq:WHRTCSLSS}
z(t)&=
\begin{cases}
\bar{C}x(t)+\bar{D}w(t)+\bar{D}_{\bar{u}}u(t)&l=1\\
\mat{c}{\bar{C}\\\bar{C}\bar{A}\\\vdots\\\bar{C}\bar{A}^{l-1}}x(t)+\mat{cccc}{\bar{D}&0&0&0\\\bar{C}\bar{B}&\bar{D}&0&\vdots\\\vdots&\ddots&\ddots&0\\\bar{C}\bar{A}^{l-2}\bar{B}&\hdots&\bar{C}\bar{B}&\bar{D}}w(t)+\mat{c}{\bar{D}^{\bar{u}}\\\bar{C}\bar{B}^{\bar{u}}\\\bar{C}\bar{A}\bar{B}^{\bar{u}}\\\vdots\\\bar{C}\bar{A}^{l-2}\bar{B}^{\bar{u}}}u(t)&l\neq 1\\
\end{cases}}for each label $l=1,...,m$. The representations for the hold strategy can be obtained with similar calculations~\cite{Seidel2023}. Observe that the state-space matrices are the same as those which appear in the context of sampled data systems~\cite{Chen1995}. This is not surprising because the underlying idea of lifting the system representation is the same. The motivation to study the derived CSLS with the family of state-space representations defined by~\eqref{eq:WHRTCSLSS} (we abbreviate this CSLS by WHRT-CSLS) is given by the following lemma.
\begin{lemma}
The WHRT control system satisfies quadratic performance described by $P=\mat{cc}{Q&S\\S\tp&R}$ if and only if the WHRT-CSLS satisfies quadratic performance with index\aln{(P^p_l)_{l=1}^m=\left(\mat{cc}{I_l\kron Q&I_l\kron S\\I_l\kron S\tp&I_l\kron R}\right)_{l=1}^{m}.}
\end{lemma}
\proofen{We first show the statement regarding asymptotic stability. Similar arguments as in~\cite{Linsenmayer2017} show that asymptotic stability of the WHRT-CSLS implies the same for the WHRT control system. For the converse, note that the state trajectory of the WHRT-CSLS is a sub-sequence of the state trajectory of the WHRT control system and hence if the latter is asymptotically stable, then the former is asymptotically stable as well. For the statement regarding performance observe that for given input and output trajectories $\bar{w}$ and $\bar{z}$ of the WHRT control system the equality\aln{
&\sum \limits_{t=0}^{\infty}\mat{c}{\bar{w}(t)\\\bar{z}(t)}\tp \mat{cc}{Q&S\\S\tp&R}\mat{c}{\bar{w}(t)\\\bar{z}(t)}\\&=\sum \limits_{\tilde{t}=0}^{\infty}\mat{c}{q_{d_\mathrm{i}(\sigma(e))}(\bar{w})(\tilde{t})\\q_{d_\mathrm{o}(\sigma(e))}(\bar{z})(\tilde{t})}\tp\mat{cc}{I_l\kron Q&I_l\kron S\\I_l\kron S\tp&I_l\kron R}\mat{c}{q_{d_\mathrm{i}(\sigma(e))}(\bar{w})(\tilde{t})\\q_{d_\mathrm{o}(\sigma(e))}(\bar{z})(\tilde{t})}\\&=\sum \limits_{\tilde{t}=0}^{\infty}\mat{c}{w(\tilde{t})\\z(\tilde{t})}\tp\mat{cc}{I_l\kron Q&I_l\kron S\\I_l\kron S\tp&I_l\kron R}\mat{c}{w(\tilde{t})\\z(\tilde{t})}}holds. We used that every admissible sequence of losses in the WHRT control system can be alternatively described by a sequence of labeled edges $e$. Together with the fact that $q_{d_\mathrm{i}(\sigma(e))}$ and $q_{d_\mathrm{o}(\sigma(e))}$ are isometric isomorphisms, the claim follows.
}
In view of this lemma, we can apply all results which we obtained in this work to design the desired controller~\eqref{eq:WHRTcon} after rewriting the WHRT control system into the WHRT-CSLS. We cannot only design a classical state-feedback controller~\eqref{eq:WHRTcon} but also a state-feedback controller which switches dependent on the past loss sequence. The latter one is a node-dependent controller and the former one is the node-independent controller from Section~\ref{sec:synthesis}. If the system matrices in~\eqref{eq:WHRTsystem} are uncertain, also the state-space realizations involved in the WHRT-CSLS are affected by uncertainties. However, we can use the results discussed in Section~\ref{sec:robustness} and~\ref{sec:synthesis} to obtain a robust controller. In this way, the presented results provide a method to design controllers for WHRT control systems which achieve robust quadratic performance. Moreover, a particular structure of the uncertainty can be taken into account during controller synthesis. For robustness analysis or synthesis, it is important to have LFRs of the uncertain state-space representations assigned to the different labels. If such a representation is given for the original WHRT control system, then a natural way is to use LFT calculus (e.g. see \cite{Zhou1996}) to obtain LFRs for the representations which are assigned to the different labels.

We remark that for the special case of nominal $\ell^2$-performance the obtained results are the same as those which we presented in~\cite{Seidel2023}. However, the benefit of the approach presented in this chapter is twofold. First, it is possible to consider arbitrary quadratic performance criteria for WHRT control systems and this also in the case of uncertainties. Second, the results of this chapter are based on simple mechanisms, provided by dissipativity, which allow a system theoretic interpretation of them. This is not directly possible with the arguments provided in~\cite{Seidel2023}.
\section{NUMERICAL EXAMPLE}\label{sec:numerical}
\begin{figure}
	\centering
	\begin{tikzpicture}
		\tikzset{vertexFill/.style = {shape=circle,fill=gray!50!white,draw,minimum size=1.5em,inner sep=0.3em},thick}
		\tikzset{edge/.style = {->,> = latex',thick}}
		\node[vertexFill] (v1) at (0,0) {\Large$v_1$};
		\node[vertexFill] (v2) at (2,0) {\Large$v_2$};
		\draw[edge] (v1) to[loop above] node[above, pos=0.5] {\large$1$} (v1);
		\draw[edge] (v1) to[bend right] node[below, pos=0.5] {\large$2$} (v2);
		\draw[edge] (v2) to[bend right] node[above, pos=0.5] {\large$1$} (v1);
	\end{tikzpicture}
	\caption{The constraining graph for the CSLS, c.f.~Fig.~\ref{fig:graph-construction}.}
	\label{fig:graph-example}
\end{figure}
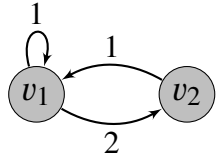

As a numerical example, we consider the problem of synthesizing a state-feedback controller for a control system with packet dropouts as discussed in Section~\ref{sec:WHRT}. The system is defined by\alnl{
\nonumber\bar{x}(t+1)&=\mat{cc}{0&1\\1&1}\bar{x}(t)+\mat{c}{1\\1}\bar{w}(t)+\mat{c}{0\\1+\delta}\bar{u}(t)\\
\label{eq:ex_sys}\bar{z}(t)&=\mat{cc}{1&1}\bar{x}(t)+\bar{w}(t)+(1+\delta)\bar{u}(t)} with the uncertainty $\delta\in\mathbb{C}$ which satisfies $\norm{\delta}\leq 1$. For example, the uncertainty parameter $\delta$ can capture an uncertainty in the model of the electronic device which receives the control signal and converts it into the system input, e.g., a force. A slightly modified variant of the described system is used as an example system in the literature for WHRT control systems~\cite{Blind2015,Linsenmayer2017}. For the losses, we assume that in every time window of length three, there are at least two time instants where the control information is not lost and that the zero strategy is applied. The objective is to find a constant state-feedback controller which minimizes $\gamma>0$ such that\aln{
\norm{z}_{\ell^2}\leq\gamma^2\norm{w}_{\ell^2}} for all input signals $w\in \ell^2$ and $\bar{x}(0)=0$. By following the steps discussed in Section~\ref{sec:WHRT}, the problem can be reformulated into a non-switching state-feedback design problem for the CSLS defined by the constraining graph\aln{(\V,\E)=(\{1,2\},\{(1,1,1), (1,2,2), (2,1,1)\})}where the system which is mapped to label one is given by~\eqref{eq:ex_sys} and for the label two, it is given by~\alnl{\bar{x}(t+1)&=\mat{cc}{1&1\\1&2}\bar{x}(t)+\mat{cc}{1&1\\2&1}\bar{w}(t)+\mat{c}{1+\delta\\1+\delta}\bar{u}(t)\nonumber\\
\label{eq:ex_sys2}\bar{z}(t)&=\mat{cc}{1&1\\1&2}\bar{x}(t)+\mat{cc}{1&0\\2&1}\bar{w}(t)+\mat{c}{1+\delta\\1+\delta}\bar{u}(t).}
The constraining graph is also illustrated in Figure~\ref{fig:graph-example}.
Observe that the state-space representation which is mapped to label two has a different input/output dimension than the representation mapped to label one.
We can use the results of Section~\ref{sec:synthesis} to design a non-switching state-feedback controller for the case $\delta=0$. During synthesis, we can guarantee that $\gamma\leq 3.6707$. This is also possible with the results presented in~\cite{Seidel2023}. However, we have no guarantees about whether the controller stabilizes the system for all possible values of $\delta$ and we do not know which upper bound of $\gamma$ can be guaranteed in these cases. In order to analyze the controlled interconnection from this perspective, we can use the results from Section~\ref{sec:robustness}. It is a simple calculation to obtain LFRs for the representations~\eqref{eq:ex_sys} and~\eqref{eq:ex_sys2} with the uncertainty\aln{w_u(t)=\Delta z_u(t),\quad \Delta\in\DeltaB=\{\tilde{\Delta}\in\mathbb{C}:\norm{\tilde{\Delta}}\leq 1\}}where, similar to Subsection~\ref{sec:robustness}, $z_u$ denotes the input of the uncertainty and $w_u$ its output. Note that in this special case, the structure of the uncertainty is the same for the state-space representation which corresponds to label one and for that which corresponds to label two. If we define\aln{\P^{\Delta}_l=\left\{\mat{cc}{-a&0\\0&a}:a\geq0\right\},\quad\text{for }l=1,2,}then it is clear that~\eqref{eq:SepUn} holds. We can now use the results from Subsection~\ref{sec:robustness} to analyze the controlled interconnection with respect to robust $\ell^2$-performance. We obtain that the controller stabilizes the system for each admissible $\delta$ and the closed-loop satisfies robustly a guaranteed bound of $\gamma\leq 6.8472$. At this point, it is crucial to remind that for synthesizing a non-switching controller we have to set $G_1=G_2$ which we do not have to do for analyzing the closed-loop. For the sake of completeness, we perform also a robust performance analysis with the restriction $G_1=G_2$ and obtain an upper bound $\gamma\leq 7.0049$. It is not surprising that the guaranteed bound of robust performance without a constraint on $G_1$ and $G_2$ is lower than that with the constraint $G_1=G_2$. Moreover, it is also not surprising that the guaranteed bound for robust performance is higher than that of the nominal case.
With the proposed results we cannot only analyze a closed-loop with respect to robust performance but we can also perform synthesis. Performing a robust performance controller synthesis, we can guarantee a robust bound of $\gamma\leq 6.7094$ during synthesis. This bound is lower than the bound obtained by a robustness analysis of the nominal controller. Analyzing the closed-loop with the robust controller yields the bound $\gamma\leq 6.2371$ which is even significantly lower. We can also consider guaranteed bounds for fixed values of $\delta$. For example, for the values $\{-0.2,0.1,0,0.1,0.2\}$ the corresponding bounds are displayed in Table~\ref{tab:Bounds}. 
			\begin{table}
				\centering
					\caption{Guaranteed bounds for the $\ell^2$-gain for different values of $\delta$}
					\label{tab:Bounds}
					\begin{tabular}{c|c|c|c|c|c}
								&\multicolumn{5}{c}{value of $\delta$}\\
								&-0.2&-0.1&0&0.1&0.2\\\hline
						    nominal controller&3.4358&3.1612&3.4861&3.9482&5.0226\\
								robust controller&3.7707&3.0706&3.2543&3.7670&4.1655
					\end{tabular}
			\end{table}
As expected, the guaranteed bounds for the different values of $\delta$ are lower than those which we obtained by performing a robust performance analysis. The significant difference can be explained by the fact that we perform the robust performance analysis for a much broader class of uncertainties than those which we consider in Table~\ref{tab:Bounds} and that we are looking for a set of certificates which certifies the bound on $\gamma$ for all possible uncertainties during robust performance analysis. In contrast, to obtain the values in Table~\ref{tab:Bounds} it suffices to find a set of certificates for each value of $\delta$ independently. It is interesting to observe that in the nominal case, the guaranteed bound for the robust controller is lower than that for the nominal controller.

We do not want to interpret the obtained numerical insights in more detail because such a discussion is highly problem-related, but we want to emphasize that the results presented in this work allow us to generate such insights. This was not possible with previous work.
\section{CONCLUSION}
In this chapter, we proposed a framework for analyzing and designing state-feedback controllers for CSLSs with respect to robust quadratic performance criteria. We considered CSLSs which are defined by a labeled graph and a finite set of linear state-space representations. The state-space representations are allowed to have different input and output dimensions and can be affected by possible time-varying and nonlinear uncertainties. It was only assumed that the uncertain state-space representations can be rewritten in LFRs. We demonstrated how the problem of verifying robust quadratic performance can be cast into the problem of verifying dissipativity with a node-dependent quadratic storage function and a label-dependent quadratic supply rate. We showed that verifying such dissipativity properties is possible with a convex optimization problem constrained with LMIs. Moreover, we showed how the presented framework can be used to design state-feedback controllers for CSLSs such that the controlled interconnection is dissipative. Therefore, the presented methods can be applied to design state-feedback controllers for CSLSs such that the controlled interconnection is robustly stable and satisfies robustly a desired quadratic performance criterion. The design procedure can be applied to design switching and non-switching controllers. We demonstrated the flexibility of the presented framework on the problem of finding bounds on the energy-to-peak gain of a CSLS.

Moreover, the proposed framework was applied to the problem of designing robust state-feedback controllers for WHRT control systems such that the closed-loop systems satisfy robustly given quadratic performance criteria. In this way, the proposed framework can also be used to investigate the impact of different WHRT constraints on the achievable performance of WHRT control systems. In applications of WHRT control systems, e.g., networked control or real-time control, this possibility can be exploited to gain deep insights into the considered problem, and to design controllers with guaranteed performance.

The presented framework thus provides methods to guarantee robust performance and stability of cyber-physical systems such as NCSs and control systems in a real-time settings. It therefore can be used as a tool to improve existing control methods for such systems, yielding an increase of control performance and robustness. The provided theoretic guarantees moreover allow for operation in safety-critical control tasks.

In future research, the proposed framework can be extended to the case where the uncertainties admit a description by integral quadratic constraints. Another interesting open research question is the design of dynamic output feedback controllers. While such controllers can be easily constructed when they are allowed to be dependent on the current label, it is interesting and important to find controller design methods for the cases where the controller should not depend on the current label or should be a non-switching controller.
\section*{Acknowledgment}
	We acknowledge the support by the Stuttgart Center for Simulation Science (SimTech). F. Allgöwer thanks the German Research Foundation (DFG) for support of this work within grant AL 316/13-2 and within the German Excellence Strategy under grant EXC-2075 - 285825138; 390740016.
\printbibliography
\end{document}